\documentclass[12pt]{article}
    \usepackage[margin=1in]{geometry}
    \usepackage[utf8]{inputenc}
    \usepackage{amsmath,amssymb,amsfonts,amsthm}
    \usepackage{array,tabularx,lscape}
    \usepackage{mathtools,rotating}
    \usepackage[symbol]{footmisc}
    \usepackage{physics}
    \usepackage{caption}
    \usepackage{graphicx}
    \usepackage[ruled,vlined]{algorithm2e}
    \usepackage{siunitx}
    \usepackage[colorinlistoftodos]{todonotes}
    \usepackage{hyperref}
\theoremstyle{plain}
\newtheorem{theorem}{Theorem}
\theoremstyle{definition}
\newtheorem{definition}[theorem]{Definition}
\newtheorem{example}[theorem]{Example}
\renewcommand{\thefootnote}{\fnsymbol{footnote}}
\SetKwInput{KwInput}{Input}
\SetKwInput{KwOutput}{Output}

\graphicspath{{./images/}}
\begin{document}
% To decrease the distance between the captions and the figures/tables

% The title page
\begin{titlepage}
   \begin{center}
       \vspace{0.5cm}

       \LARGE\textbf{Symmetries of systems of first order ODEs:\\Symbolic symmetry computations, mechanistic model construction and applications in biology}
       
        \normalsize
            
        \vspace{1.5cm}
        \setcounter{footnote}{1}
        {\large Johannes Borgqvist\footnote{\label{Oxford}Wolfson Centre for Mathematical Biology, Mathematical Institute, University of Oxford, United Kingdom}\footnote{\label{Linacre} Linacre College, University of Oxford, United Kingdom}, Fredrik Ohlsson\footnote{\label{Umeå}Department of Mathematics and Mathematical Statistics, Umeå University, Sweden} and Ruth E. Baker\footref{Oxford}}

        \vspace{2cm}
        \begin{abstract}
        We discuss the role and merits of symmetry methods for the analysis of biological systems. In particular, we consider systems of first order ordinary differential equations and provide a comprehensive review of the geometrical foundations pertinent to symmetries of such systems. Subsequently, we present an algorithm for finding infinitesimal generators of symmetries for systems with rational reaction terms, and an open-source implementation of the algorithm using symbolic computations. We discuss two complementary perspectives on symmetries in mechanistic modelling; as tools for the analysis of a given model or as a geometrical principle for incorporating biological properties in the construction of new models. Through numerous examples of relevance to modelling in biology we demonstrate the different uses of symmetry methods, and also discuss how to infer symmetries from experimental data.
        \end{abstract}
       
       \bigskip
            
   \end{center}
   \end{titlepage}
   
\setcounter{footnote}{0}
\renewcommand{\thefootnote}{\arabic{footnote}}

% Setup equation numbering
\numberwithin{equation}{section}

% The introduction
\section{Introduction}
\label{sec:introduction}
Mathematical modelling now constitutes an integral part of the biological and biomedical sciences, with models and experiments used in combination to better understand complex biological mechanisms, guide treatments and direct public policy. However, the process of model construction remains a fundamental problem in the field. Biological systems are enormously complex and a model is, by definition, a simplified representation of reality. This means that wide-ranging assumptions need to be made, both to reduce the complexity of a mechanism to a point where models can give useful insights, and to bridge knowledge gaps where mechanisms are incompletely understood.

Consequently, it is often possible to construct multiple models of the same system based on mutually exclusive biological mechanisms, manifested in different mathematical model structures. The resulting \textit{model selection} problem is typically difficult to resolve conclusively using experimental data. Even when it is possible to select one model out of a set of candidates based on agreement with experiments, model selection is merely a relative comparison among the candidates; there is no guarantee that another model that describes the system more adequately does not exist. 

Recently, modelling efforts have incorporated statistical learning methods, e.g.~artificial neural networks, to derive models directly from experimental data without imposing restrictive assumptions on model structures~\cite{wang2016,alquraishi2019}. However, this approach to mathematical biology is still in its infancy, and the resulting statistical models are typically plagued by low interpretability. Inspired by mathematical physics, we propose that another theoretical perspective on the problem of constructing and analysing mechanistic models can provide important insights and powerful methods. This approach is based on the powerful machinery of differential geometry in general, and on the concept of symmetries of differential equations in particular, and it offers a complementary theoretical approach to the existing methods.

Simply put, a symmetry of an object is an operation that leaves the object invariant. For example, the unit circle is unaffected by rotations around the origin. Mathematically, a symmetry is a transformation that preserves some property of the object. In the example of unit circle, the object under consideration is the equation $x^2+y^2=1$, the transformation is
\begin{equation}
\Gamma_{\epsilon}(x,y) = (\cos(\epsilon)x - \sin(\epsilon)y, \, \sin(\epsilon)x + \cos(\epsilon)y),
\end{equation}
where $\epsilon \in \mathbb{R}$ is the angle of rotation, and the object preserved is the space of solutions\footnote{In this case, of course, the space of solutions consists of a single curve in the $(x,y)$-plane, and the solution itself is also invariant under the transformation $\Gamma_{\epsilon}$.}.

The notion of symmetries can be extended to differential equations as transformations, acting on both independent and dependent variables, that map one solution to another. Here, we exclusively consider systems of first order \textit{ordinary differential equations} (ODEs) as these are common in mathematical biology. In this context, the independent variable is often time and denoted by $t$, and the dependent variables are the states which we will denote by $y=(y_1,\ldots,y_k)$, corresponding to, for example, the concentrations of a set of proteins or the sizes of a set of populations at time $t$. The time derivatives in $\dd y/\dd t$ are equal to some (often non-linear) functions $\omega_i(t,y)$, $i=1,\ldots,k$, referred to as the reaction terms. Moreover, we use the term \textit{model} for systems of first order ODEs, and the qualification \textit{mechanistic model} refers to the fact that various biological assumptions on the reaction or growth rate are mathematically encoded in the reaction terms $\omega_i(t,y)$. 

% --------------------------------------------------------------------

\begin{figure}[htbp!]
\begin{center}
\includegraphics[width=0.8\textwidth]{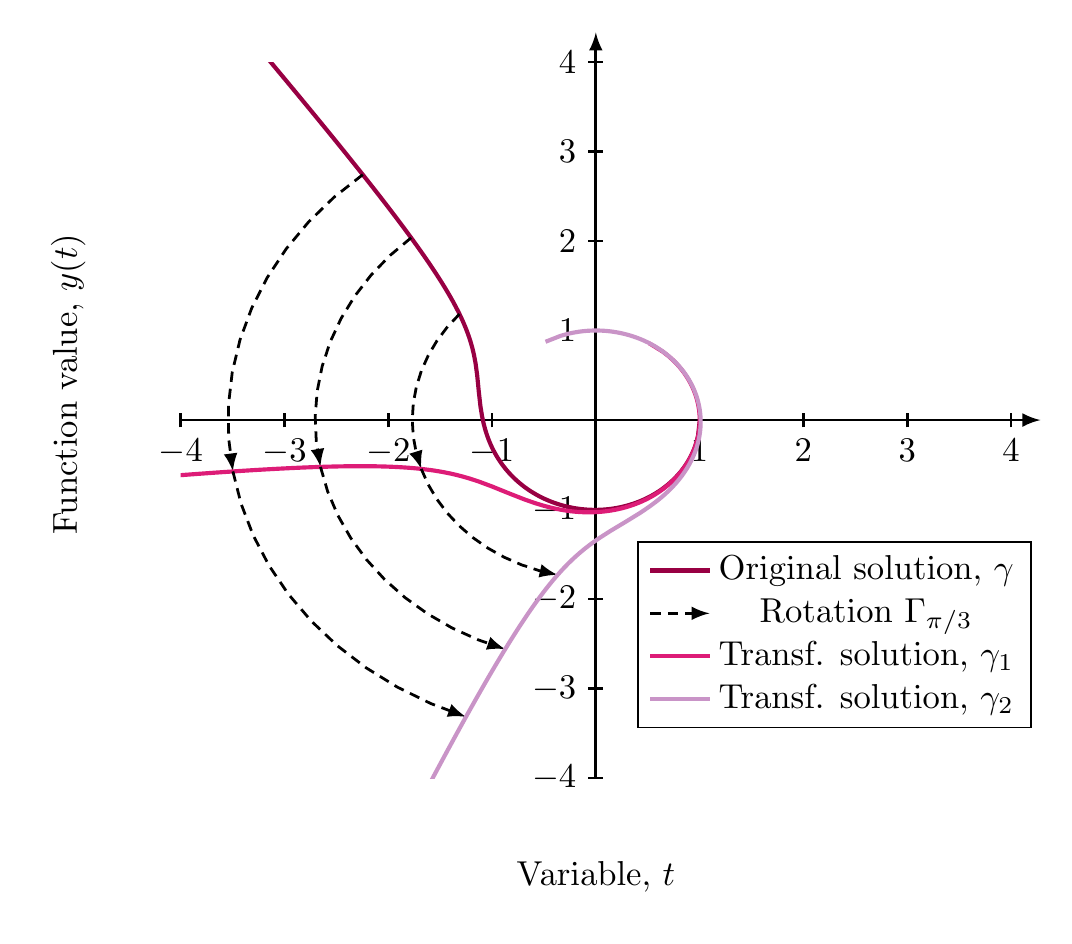}
\caption{An illustration of the action of the anticlockwise rotation $\Gamma_{\epsilon}$ in Equation~\eqref{eq:intro_symmetry_example_transf} with an angle of $\epsilon=\pi/3$ on solutions to the ODE in Equation~\eqref{eq:intro_symmetry_example_equation}. Repeated rotations map the solution $\gamma$ to two other solutions where $\gamma_1 = \Gamma_{\pi/3} \gamma$ and $\gamma_2 = \Gamma_{\pi/3} \gamma_1 = \Gamma_{\pi/3}^2 \gamma$.}
\label{fig:intro_symmetry_example}
\end{center}
\end{figure}

\begin{example}
Consider the ODE
\begin{equation}
\label{eq:intro_symmetry_example_equation}
\frac{\dd y}{\dd t} = \omega(t,y)=\frac{(y^3+t^2y-y-t)}{(ty^2+t^3+y-t)},
\end{equation}
and the transformation
\begin{equation}
\label{eq:intro_symmetry_example_transf}
\Gamma_{\epsilon}(t,y) = (\cos(\epsilon)t-\sin(\epsilon)y,\,\sin(\epsilon)t+\cos(\epsilon)y) ,
\end{equation}
which amounts to an anticlockwise rotation in the $(t,y)$-plane by an angle $\epsilon \in \mathbb{R}$. The transformation $\Gamma_\epsilon$ in Equation~\eqref{eq:intro_symmetry_example_transf} maps solutions of Equation~\eqref{eq:intro_symmetry_example_equation} to other solutions, as illustrated in Figure~\ref{fig:intro_symmetry_example}. Consequently, the rotation transformation $\Gamma_\epsilon$ in Equation~\eqref{eq:intro_symmetry_example_transf} preserves the space of solutions and constitutes a symmetry of Equation~\eqref{eq:intro_symmetry_example_equation}~\cite{hydon2000symmetry}.
\hfill $\square$
\end{example}

% --------------------------------------------------------------------

% --------------------------------------------------------------------

The appeal of symmetries in modelling is that they encode properties of the mechanisms governing the underlying system, and can be used to find analytical solutions, derive conservation laws and even construct models from first principles~\cite{hydon2000symmetry,bluman1989symmetries,bluman2010applications, olver2000applications,stephani1989differential}. In fundamental physics, geometrical formulations have been used with great success to construct, analyse and validate models~\cite{Gross1996}. Existing applications of symmetry methods in mathematical biology, reviewed in \cite{golubitsky2015symmetry}, include finding analytical solutions to reaction--diffusion models~\cite{cherniha2000,davydovych2018,cherniha2017nonlinear}, conducting model selection based on using symmetry transformations to infer model structure~\cite{ohlsson2020symmetry}, and performing identifiability analysis of systems of first order ODEs~\cite{yates2009,castro2020,massonis2020finding}. In the last case, an algorithm for finding a certain class of symmetries of first order dynamical systems has been implemented~\cite{merkt2015higher}.

Unfortunately, the scale and complexity of many models in mathematical biology renders a brute force application of symmetry methods impractical. However, a geometrical formulation of the constituent systems offers a complementary approach to the formidable problem of first-principle model construction and can provide novel biological insights. In addition, the interpretation of symmetries as mathematical operators that encode physical or biological properties provides a means of assembling simple constituents together to give models of complex systems that incorporate fundamental biological principles into the model structure. 

Here, we argue the merits of using symmetries as a fundamental principle for analysing and constructing mechanistic models of biological systems where biophysical properties are incorporated in the very structure of the models. In order to automate the calculations of symmetries, we describe an algorithm for finding a large and common class of symmetries, extending the scope of~\cite{merkt2015higher}, for models consisting of systems of first order ODEs with polynomial (e.g.~mass action kinetics) or rational (e.g.~Hill functions) reaction terms. These types of models are frequently used in the modelling of biological systems and serve as an ideal starting point for elucidating the role of symmetry methods in mathematical biology. Subsequently, we present an open-source implementation of this algorithm based on the symbolic solver SymPy~\cite{meurer2017sympy} and, using this implementation, we find the symmetries of some common models in mathematical biology. By interpreting these symmetries, we show how well-known properties of these models emerge through their symmetries. Thereafter, we reverse the theoretical analysis so that instead of finding the symmetries of well-known models we demonstrate how symmetries can be used to construct models and discuss strategies for inferring the mathematical structure required to capture biological mechanisms underlying a phenomenon from experimental observations of it. 

As the analysis of ODEs using symmetry methods and other techniques based on differential geometry are non-standard in mathematical biology, we will initially summarise the mathematical framework for analysing and constructing mechanistic models using symmetries. To this end, we will begin by providing an overview of aspects of differential geometry pertaining to symmetries of first order ODEs in Section~\ref{sec:geometry}. In Section~\ref{sec:algorithm}, we present an algorithm for finding symmetries using symbolic calculations and then discuss our computational implementation before considering some examples of its application to systems of ODEs. In Section~\ref{sec:bioexamples}, we interpret the biological meaning of the symmetries of some well-known models calculated using our implementation of the algorithm. Lastly, we reverse the focus in order to present a symmetry-based methodology for the construction of mechanistic models in Section~\ref{sec:model_construction}, before providing a discussion on data-driven symmetry discovery in Section~\ref{sec:symmetry_discovery} and our work in general in Section~\ref{sec:discussion}.

% Mathematical Theory
\section{Symmetries of ODEs: a geometrical perspective}
\label{sec:geometry}
In this section we present the geometrical framework of jet spaces (and bundles) which appears in the analysis of symmetries of differential equations. We motivate the construction of this framework using the familiar notion of systems of differential equations and their solutions, and we emphasise the geometrical formulation of these objects as well as transformations acting on them. The presentation is based on differential geometry in general, and the concepts of manifolds, fibre bundles and Lie groups in particular. Moreover, the aim here is to present the geometrical foundations of symmetries for first order ODEs in particular but this theory generalises to any type of differential equation. For the interested reader, there are many excellent introductory texts~\cite{hydon2000symmetry,olver2000applications,olver2008equivalence,lang2001,nakahara2003} available which provide a more in-depth overview of these topics and their applications to differential equations.

% ---------------------------------------------------------------------

\subsection{Differential equations and jet space}\label{sec:DEs_and_JS}

We consider systems of ODEs in one independent variable, $t$, and $k$ dependent variables $y_1,\ldots,y_k$ given by
\begin{equation}
\label{eqn:ode_system}
\frac{\dd y_i}{\dd t} = \omega_i(t,y_1,\ldots,y_k), \quad \quad i = 1,\ldots,k.
\end{equation}
In order to make the notion of symmetries of the system of ODEs in Equation~\eqref{eqn:ode_system} precise, and develop the tools required to study them, we make use of a geometrical formulation where the variables $t$ and $y_1,\ldots,y_k$ are considered as local coordinates on $T \simeq \mathbb{R}$ and $U \simeq \mathbb{R}^k$, respectively, and together parametrise a manifold of fundamental importance.

% ---------------------------------------------------------------------

\begin{definition}
The \emph{total space} is given by the direct product $E = T \times U$. The natural projection $\pi$ to the first factor equips $E$ with the structure of a fibre bundle $\pi: E \rightarrow T$. A point in $E$ is denoted by $(t,y)$ where $y = (y_1,\ldots,y_k) \in U$.
\hfill $\square$
\end{definition}

% ---------------------------------------------------------------------

A smooth function $y = f(t)$, such that  
\begin{equation}
\begin{array}{cccl}
f: & T & \to & U,\\
& t & \mapsto & (f_1(t),\ldots,f_k(t)),
\end{array}
\end{equation}
defines a (local) section $\gamma_f$ of the bundle $E$ through its graph
\begin{equation}
\begin{array}{cccl}
\gamma_f: & T & \rightarrow & E,\\
& t & \mapsto & (t,f_1(t),\ldots,f_k(t)).
\end{array}
\end{equation}
The fact that Equation~\eqref{eqn:ode_system} describes a system of ODEs implies that the geometrical formulation must also include the derivatives $y'_1,\ldots,y'_k$ of the dependent variables with respect to the independent variable $t$. To this end, we introduce an extension of the total space $E$ by considering the space $U_1 \simeq \mathbb{R}^k$ parametrized by $y'_1,\ldots,y'_k$. 

% ---------------------------------------------------------------------

\begin{definition}
The first \emph{jet space} associated to $E = T \times U$ is the product space $J^{(1)} = T \times U \times U_1 = T \times U^{(1)}$ where $U^{(1)} = U \times U_1$. The natural projection $\pi^{(1)}$ to the factor $T$ equips $J^{(1)}$ with the structure of a fibre bundle called the first \emph{jet bundle} $\pi^{(1)} : J^{(1)} \to T$. A point in $J^{(1)}$ is denoted by $(t,y^{(1)})$ where $y^{(1)} = (y_1,\ldots,y_k,y'_1,\ldots,y'_k) \in U^{(1)}$.
\hfill $\square$
\end{definition}

% ---------------------------------------------------------------------

Any function $y=f(t)$ and its corresponding local sections $\gamma_f$ can be extended, or \emph{prolonged}, to the jet space $J^{(1)}$ through computation of the corresponding derivatives.

% ---------------------------------------------------------------------

\begin{definition}
Let $f : T \to U$ be a function and $\gamma_f : T \to E$ be the corresponding section. The \emph{prolonged function} $f^{(1)}$ and \emph{prolonged section} $\gamma_f^{(1)}$ are induced by lifting to $U^{(1)}$ and $J^{(1)}$, respectively,
\begin{equation}
\begin{array}{cccl}
f^{(1)}: & T & \to & U^{(1)},\\
& t & \mapsto & \left( f_1(t),\ldots,f_k(t),\frac{\text{d}}{\text{d}t}f_1(t),\ldots,\frac{\text{d}}{\text{d}t}f_k(t) \right),
\end{array}
\end{equation}
\begin{equation}
\begin{array}{cccl}
\gamma_f^{(1)}: & T & \to & J^{(1)},\\
& t & \mapsto & \left( t,f_1(t),\ldots,f_k(t),\frac{\text{d}}{\text{d}t}f_1(t),\ldots,\frac{\text{d}}{\text{d}t}f_k(t) \right).
\end{array}
\end{equation}
\hfill $\square$
\end{definition}

% ---------------------------------------------------------------------

The geometrical interpretation of the system of ODEs given in Equation~\eqref{eqn:ode_system} is obtained through the smooth map $\Delta : J^{(1)} \to \mathbb{R}^k$ with components
\begin{equation}
\label{eqn:delta}
\Delta_i(t,y^{(1)}) = y'_i-\omega_i(t,y), \quad \quad i=1,\ldots,k,
\end{equation}
which defines a subvariety of the jet space $J^{(1)}$ through
\begin{equation}
S_{\Delta} = \left\{ (t,y^{(1)}) \, | \, \Delta(t,y^{(1)})=0 \right\} \subset J^{(1)}.
\end{equation}
We will sometimes refer to the system of ODEs given in Equation~(\ref{eqn:ode_system}) simply by the corresponding function $\Delta$. If the map $\Delta$ in Equation~\eqref{eqn:delta} has constant rank on $J^{(1)}$ the system is said to be regular.

% ---------------------------------------------------------------------

\begin{definition}
A (local) \emph{solution} to a system $\Delta$ of ODEs is a smooth function $y = f(t)$ whose prolongation satisfies $\Delta(t,f^{(1)}(t)) = 0$ or, equivalently, whose prolonged section $\gamma^{(1)}_f$ is contained entirely in the corresponding subvariety $\gamma^{(1)}_f \subset S_{\Delta}$.
\hfill $\square$
\end{definition}

% ---------------------------------------------------------------------

Given an initial condition $y_0 = f(t_0)$ there exists a unique local solution to $\Delta$ if the reaction terms on the right-hand side of the system of ODEs in Equation~\eqref{eqn:ode_system}, i.e.~$\omega_i(t,y)$ for $i=1,\ldots,k$, are smooth functions. For the purpose of describing local solutions, two functions are clearly equivalent if their component values and first derivatives are identical. Therefore, a coordinate-independent definition of jet space $J^{(1)}$ is obtained from the space of functions on the total space $E$ by identifying all functions whose prolongations are equal.

% ---------------------------------------------------------------------

\begin{figure}[htbp!]
\begin{center}
    \includegraphics[width=0.8\textwidth]{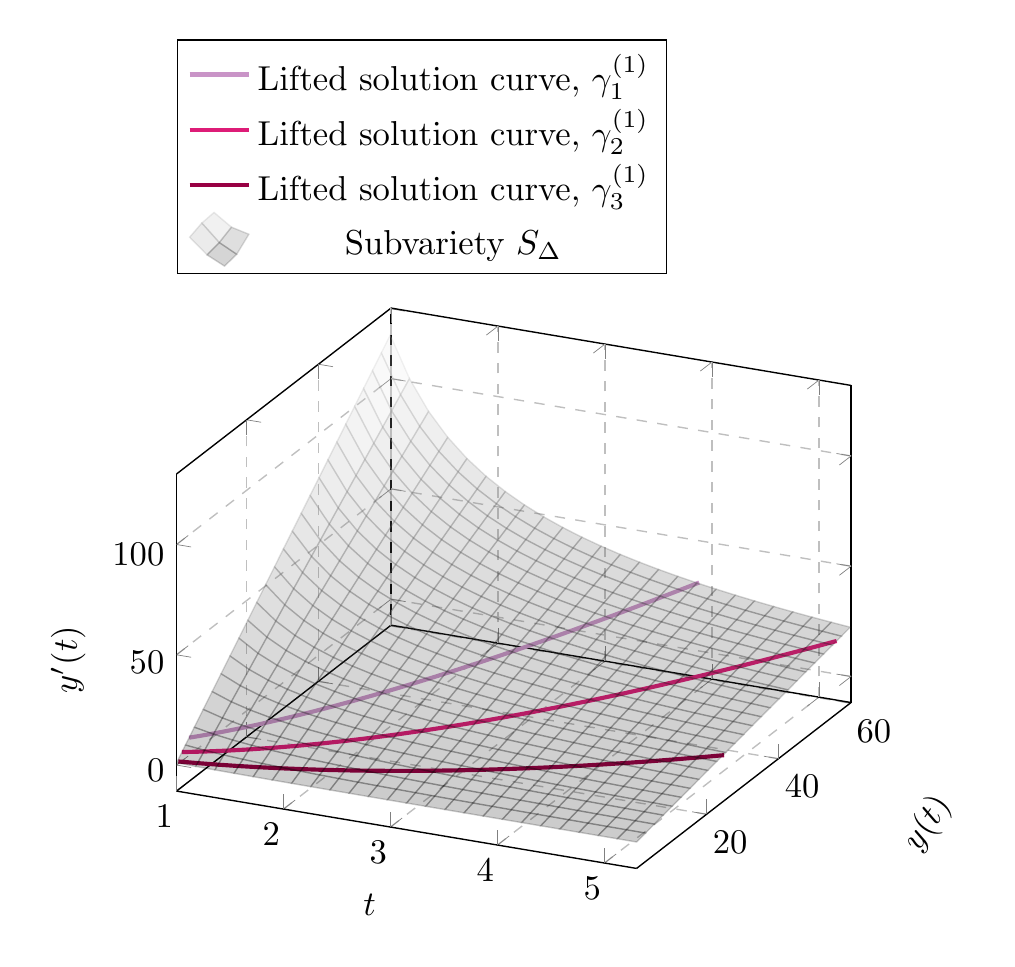}
    \caption{An illustration of three lifted solution curves to the ODE in Equation~(\ref{eqn:single_ode_example}) given by $\gamma^{(1)}_1=(t,4t^2,8t)$, $\gamma^{(1)}_2=(t,2t^2,4t)$ and $\gamma^{(1)}_3=(t,t^2,2t)$. These lifted solution curves are obtained by setting the arbitrary constant $C_1$ in  Equation~(\ref{eqn:singel_ode_solution_example}) to $C_1=4$, $C_1=2$ and $C_1=1$, respectively. Moreover, the subvariety $S_{\Delta}$ containing the set of lifted solution curves is illustrated by the grey surface.}
\label{fig:prolongations_example}
\end{center}
\end{figure}

\begin{example}
Consider the ODE given by
\begin{equation}
\label{eqn:single_ode_example}
\frac{\dd y}{\dd t} = \omega(t,y) = \frac{2y}{t}.
\end{equation}
The general (local) solution is given by $y = f(t) = C_1 t^2$ for $C_1$ an arbitrary constant. The solution curve in $E$ is given by the graph $\gamma_f = \{(t,C_1t^2)\}$ with lift
\begin{equation}
\label{eqn:singel_ode_solution_example}
    \gamma^{(1)}_f = \{(t,C_1t^2,2C_1t)\},
\end{equation}
in $J^{(1)}$. We note that the solutions are well-defined everywhere in $E$ and $J^{(1)}$, even though $\omega(t,y)$ is only smooth away from $t=0$, and extends to global solutions. The solutions, their prolongations and the subvariety $S_{\Delta}$ defined by the ODE are illustrated in Figure~\ref{fig:prolongations_example}.
\hfill $\square$
\end{example}

% ---------------------------------------------------------------------------------

% ---------------------------------------------------------------------

The following analysis of differential equations and their symmetries will frequently involve computing derivatives in jet space. Such computations are conveniently expressed in terms of the total derivative with respect to the independent variable.

% ---------------------------------------------------------------------

\begin{definition}
The \emph{total derivative} is defined as the differential operator   
\begin{equation}
D_t = \partial_t + y'_1\partial_{y_1} + \cdots + y'_k\partial_{y_k},
\label{eq:tot_derivative}
\end{equation}
In particular, for a function $F : E \to \mathbb{R}$ the total derivative is the unique function $D_tF : J^{(1)} \to \mathbb{R}$ satisfying 
\begin{equation}
D_tF(t,f^{(1)}(t)) =  \frac{\dd}{\dd t} F(t,f(t)),
\end{equation}  
for any function $y = f(t)$.
\hfill $\square$
\end{definition}

% ---------------------------------------------------------------------

\subsection{Transformation groups and invariants}

That an objects has a symmetry amounts to the statement that the object is invariant under some transformation. The objects we are concerned with here are the systems of ODEs discussed in Section~\ref{sec:DEs_and_JS}, and we will now turn our attention to the transformations acting on them. We will always consider continuous groups of transformations obtained through the action of some Lie group $G$, i.e. a smooth manifold equipped with a differentiable group structure, on the total space $E$ parametrised by the independent and dependent variables of the system \eqref{eqn:ode_system}. Subsequently, we describe the properties of these transformations and we make the notion of \textit{invarance} in the context of differential equations precise. 

In general, the action of $G$ on a smooth manifold $M$ is given by a differentiable map $\Phi : G \times M \to M$ such that
\begin{equation}
\Phi(g,x) = \Gamma_g x,
\end{equation}
where $\Gamma_g$ is a representation of $G$ acting pointwise on $M$. In order to define diffeomorphisms on $M$ compatible with the group structure of $G$, we require that the following holds
\begin{align}
\Gamma_ex &= x, \label{eq:identity}\\
\Gamma_g\left( \Gamma_hx \right) & = \Gamma_{gh} x,\label{eq:multiplication}
\end{align}
for all $g,h \in G$ and $x \in M$. Here, the identity element of the group is denoted by $e\in G$ and juxtaposition of group elements represents multiplication in $G$. The group $G$ is called a \emph{transformation group} on $M$ and for a fixed element $g \in G$ the map $\Gamma_g : M \to M$ is referred to as a \emph{point transformation}. We will usually allow transformation groups to be local, meaning that the transformation $\Phi$ is only defined for $g$ in some open neighbourhood of the identity $e$.

An \emph{orbit} of the transformation group $G$ is a subset $\mathcal{O} \subset M$ which is invariant under $G$, meaning that $\Gamma_g(\mathcal{O}) = \mathcal{O}$ for all elements $g \in G$. The action of $G$ is \emph{semi-regular} if all orbits have the same dimension, and \emph{regular} if in addition every orbit is a regular submanifold of $M$.

A local \emph{invariant} of the transformation group $G$ is a function $I : M \to \mathbb{R}$, defined on some open subset of $M$, satisfying $I (\Gamma_g x) = I(x)$ for all $g$ sufficiently close to the identity. Furthermore, if $G$ acts semi-regularly with orbits of dimension $s$ there are $\mu = \mathrm{dim}\,M - s$ functionally independent local invariants $I_1,\ldots,I_{\mu}$ at every point $x \in M$, and every other local invariant at $x$ can be written as a function of $I_1,\ldots,I_{\mu}$. If the action of $G$ is regular, the invariants can be extended globally. This fundamental connection between orbits and invariants will be of great importance when we later on apply symmetry methods in order to analyse and construct ODE models.

Specialising to the case $M=E$ of interest for differential equations we express the point transformation $\Gamma : E \to E$ as 
\begin{equation}
\label{eqn:point_transformation}
\Gamma (t,y) = \left( \hat{t}(t,y) , \hat{y}(t,y) \right),
\end{equation}
where $\hat{t}(t,y)$ and $\hat{y}(t,y)$ are smooth functions and the dependence on $g \in G$ is often left implicit. In order to make statements regarding the invariance of differential equations under a transformation group $G$, we must first consider the induced action of $G$ on functions $y=f(t)$ to describe how $G$ acts on solutions of a system $\Delta$.

% ---------------------------------------------------------------------

\begin{definition}
Let $f  : T \to E$ be a function and $\Gamma$ a local point transformation acting on $E$. In a neighbourhood of the point $(t_0,y_0) = (t_0,f(t_0))$ the image of the graph $\gamma_f$ under the transformation $\Gamma$ is the graph $\gamma_{\Gamma f}$ of the \emph{transformed function} $\Gamma f : T \to E$ in a corresponding neighbourhood of $(\hat{t}_0,\hat{y}_0) = \Gamma(t_0,y_0)$:
\begin{equation}
\label{eqn:transformation_action_graphs}
\Gamma \gamma_f = \{\Gamma (t,f(t))\} = \{(\hat{t},(\Gamma f)(\hat{t})\} = \gamma_{\Gamma f}\,.
\end{equation}
\hfill $\square$
\end{definition}

% ---------------------------------------------------------------------

Using the local action of $G$ on functions $y=f(t)$ we can now proceed to consider the induced pointwise action on the jet space $J^{(1)}$.

% ---------------------------------------------------------------------

\begin{definition}
Let $\Gamma$ be a local point transformation and let $y=f(t)$ be a representative of the point $(t_0,y^{(1)}_0) \in J^{(1)}$, i.e.~satisfying $(t_0,y^{(1)}_0)=(t_0,f^{(1)}(t_0))$. The \emph{prolonged transformation} $\Gamma^{(1)}$ is defined by
\begin{equation}
\label{eqn:prolonged_transformation}
\Gamma^{(1)} (t_0,y^{(1)}_0) = (\hat{t}_0,(\Gamma f)^{(1)}(\hat{t}_0)),
\end{equation}
where the action of $\Gamma$ on the point $(t_0,y_0) \in E$ is given by $\Gamma(t_0,y_0) = (\hat{t}_0,\hat{y}_0)$.
\hfill $\square$
\end{definition}

% ---------------------------------------------------------------------

The action of $\Gamma^{(1)}$ on the point $(t_0,y^{(1)}_0)$ amounts to the evaluation of the derivatives of the transformed function $\Gamma f$, which is clearly independent of the choice of representative function $f : E \to E$, making $\Gamma^{(1)}$ well-defined. A direct consequence of this definition is that the prolongation extends the action of $\Gamma$ on sections of $E$ in Equation~\eqref{eqn:transformation_action_graphs} to sections of $J^{(1)}$ according to
\begin{equation}
\label{eqn:transformation_action_prolonged_graph}
\Gamma^{(1)} \gamma^{(1)}_f = \gamma^{(1)}_{\Gamma f}.
\end{equation}

% ---------------------------------------------------------------------

\begin{definition}
Let $G$ be a local transformation group acting on $E$. The prolonged transformation group $G^{(1)}$ is obtained by prolonging each point transformation $\Gamma_g$ to $\Gamma^{(1)}_g$.
\hfill $\square$
\end{definition}

% ---------------------------------------------------------------------

We note that the prolongation of point transformations $\Gamma$ and transformation groups $G$ can, in general, only be obtained locally on $E$ due to the local nature of the action $\Gamma f$ on functions. However, by using the total derivative $D_t$ in Equation~\eqref{eq:tot_derivative} we can obtain an explicit expression for the prolonged transformation. More precisely, using $D_t$ we can extend the notation for the point transformation $\Gamma$ in Equation~\eqref{eqn:point_transformation} to the prolonged transformation $\Gamma^{(1)}$ according to
\begin{equation}
\label{eqn:prolonged_point_transformation}
\Gamma^{(1)}(t,y,y') = (\hat{t}(t,y),\hat{y}(t,y),\hat{y}'(t,y,y')).
\end{equation}   
Here, the prolongation is required to act trivially on the transformation of $(t,y)$ in order to reduce to $\Gamma$ upon restriction from $J^{(1)}$ to $E$. Using the chain rule, the transformed derivatives can then be expressed according to
\begin{equation}
\hat{y}'_i = \frac{\dd \hat{y}_i}{\dd \hat{t}} = \frac{D_t \hat{y}_i}{D_t \hat{t}}, \quad \quad i=1,\ldots,k.
\end{equation}
When classifying differential equations as invariant under some symmetry group, invariant functions on the jet space $J^{(1)}$ play a pivotal role.

% ---------------------------------------------------------------------

\begin{definition}
Let $G$ be a transformation group acting on $E$. A first order \emph{differential invariant} for $G$ is a function $I : J^{(1)} \to \mathbb{R}$ satisfying
\begin{equation}
\label{eqn:differential_invariant}
I(\Gamma^{(1)}_g (t,y^{(1)})) = I ( t,y^{(1)} ),
\end{equation}
for every $g \in G$ and $(t,y^{(1)}) \in J^{(n)}$ where the prolonged transformation is well-defined.
\hfill $\square$
\end{definition}

% ---------------------------------------------------------------------

According to the definition, ordinary invariants of the action of $G$ on $E$ are included as a subset of first order differential invariants. In order to understand the structure of differential invariants, we must consider the orbits of the prolonged group $G^{(1)}$. 

We denote the maximal orbit dimensions of $G$ and $G^{(1)}$ by $s_0$ and $s_1$, respectively. Since every orbit of $G^{(1)}$ restricts to an orbit of $G$, the orbit dimension is non-decreasing under prolongation. Furthermore, the orbit dimension is bounded from above by the dimension of the transformation group (which is unaffected by the prolongation), and thus we have that
\begin{equation}
s_0 \leq s_1 \leq \mathrm{dim}\,G.
\end{equation}
Using this result, we can deduce the number of functionally independent differential invariants of a transformation group $G$. To avoid singularities in the action of $G^{(1)}$, we restrict attention to the open subset $\Omega^{(1)} \subset J^{(1)}$ consisting of points belonging to orbits of maximal dimension, so that the action of $G^{(1)}$ on $\Omega^{(1)}$ is semi-regular with orbit dimension $s_1$. According to the general result above, on $\Omega^{(1)}$ there are then
\begin{equation}
\label{eqn:functionally_independent_differential_invariants}
\mu_1 = \mathrm{dim}\,J^{(1)} - s_1,
\end{equation}
functionally independent first order differential invariants $I_1,\ldots,I_{\mu_1}$ on $\Omega^{(1)}$ and, crucially, every invariant on $\Omega^{(1)}$ can be expressed as a function of $I_1,\ldots,I_{\mu_1}$.

% ---------------------------------------------------------------------

\subsection{Infinitesimal generators and invariance}

Having established the geometrical foundation of transformation groups acting on the jet space $J^{(1)}$, we now turn to the equivalent infinitesimal description. The ability to recover (the connected component of) a Lie group $G$ from its Lie algebra $\mathfrak{g}$ is arguably the most important property in practice for the study of invariance under symmetry groups, since it allows all computations to be linearised and performed infinitesimally. In particular, the Lie algebra $\mathfrak{g}$ is the vector space of right-invariant vector fields on $G$ which closes to an algebra under the Lie bracket\footnote{Here, we slightly abuse the notation for composition of vector fields on $G$.} $[v_1,v_2] = v_1 v_2 - v_2 v_1$ where $v_1,v_2 \in \mathfrak{g}$.

The connection between the algebra $\mathfrak{g}$ and the group $G$ is provided by the fact that a vector field $v$ on $G$ defines a unique integral curve, or \textit{flow}, through each point $g \in G$. We denote the flow  $\exp(\epsilon v)g$, where $\epsilon \in \mathbb{R}$ parametrises the curve, implying that the vector field $v\in\mathfrak{g}$ that \textit{generates} the flow is recovered as
\begin{equation}
%\label{eqn:generating_vector_field}
v \vert_g = \left. \frac{\dd}{\dd\epsilon} \left( \exp (\epsilon v) g\right) \right\vert_{\epsilon=0}.
\end{equation}
In particular, the flow through the identity $e\in G$ defines a 1-parameter subgroup of $G$, denoted $\exp(\epsilon v)$, generated by $v\in\mathfrak{g}$\footnote{The explicit connection between $\mathfrak{g}$ and $G$ is provided by the exponential map $\exp: \mathfrak{g} \to G ,\, v \mapsto \exp(v)$ obtained by evaluating the flow at $\epsilon=1$.}.

The Lie algebra of a transformation group $G$ acting on a manifold $M$ induces a Lie algebra of vector fields on $M$. Let $v \in \mathfrak{g}$ be the generator of a 1-parameter subgroup $\exp(\epsilon v)$ of a transformation group $G$ acting on the manifold $M$. The corresponding \emph{infinitesimal generator} $X(v)$ of transformations is the unique vector field on $M$ that generates the flow $\exp(\epsilon X)x$ coinciding with the action of $\exp(\epsilon v)$. In particular, this means that at every point $x \in M$  
\begin{equation}
\label{eqn:generating_vector_field}
X(v) \vert_x = \left. \frac{\dd}{\dd\epsilon} \left( \Gamma_{\exp (\epsilon v)} x\right) \right\vert_{\epsilon=0},
\end{equation}
and that $X$ provides the tangent vector to the action of the 1-parameter subgroup.
 
The induced generating vector fields form a Lie algebra $\mathfrak{g}_M$ of vector fields that is isomorphic\footnote{Under the very mild assumption that transformation group $G$ has no global isometries.} to $\mathfrak{g}$, which allows us to locally recover the action of the 1-parameter subgroup from the infinitesimal generator through the exponential map $\exp(\epsilon X) $. In what follows, we will usually leave the element $v \in \mathfrak{g}$ implicit and simply refer to the infinitesimal generator $X$. 

The first application of the infinitesimal description of a transformation group $G$ in terms the generating vector fields $X$ on $M$ is the computation of invariants of $G$. If $G$ is connected\footnote{If $G$ is not connected, the corresponding result holds on the connected component of $G$.} a function $I : M \to \mathbb{R}$ is an invariant of $G$ if and only if
\begin{equation}
\label{eqn:infinitesimal_invariance}
X(I) = 0, \quad \quad \forall \, X \in \mathfrak{g}_M.
\end{equation}  
Since the infinitesimal generators $X \in \mathfrak{g}_M$ are differential operators, the condition given in Equation~\eqref{eqn:infinitesimal_invariance} amounts to a homogeneous system of $\mathrm{dim}\,G$ differential equations.

We then consider a vector field $X$ generating the action on $E$ of a 1-parameter subgroup of $G$ and denote the transformation corresponding to $\exp(\epsilon X)$ by
\begin{equation}
\Gamma_{\epsilon} (t,y) = (\hat{t},\hat{y})  = (\exp(\epsilon X)t, \,\exp(\epsilon X)y),
\end{equation}
where we have introduced the convention to label the transformation by the parameter $\epsilon$ rather than the full group element. Furthermore, we introduce the following notation for the local components of the generating vector field
\begin{equation}
X = \xi(t,y)\partial_t + \eta_1(t,y)\partial_{y_1} + \cdots + \eta_k(t,y)\partial_{y_k},
\end{equation}
where the individual components are given by the transformation
\begin{equation}
\xi(t,y) = \left. \frac{\dd\hat{t}}{\dd\epsilon} \right\vert_{\epsilon=0}, \quad \quad \eta_i(t,y) = \left. \frac{\dd\hat{y}_i}{\dd\epsilon} \right\vert_{\epsilon=0}, \quad \quad i=1,\ldots,k.
\end{equation}

The action of the prolongation of the 1-parameter transformation group $\Gamma_{\epsilon} = \exp(\epsilon X)$ on jet space $J^{(1)}$ can also be described infinitesimally, by the prolongation of the infinitesimal generator $X$ itself.

% ---------------------------------------------------------------------

\begin{definition}
Let $X$ be a vector field on $E$ generating the 1-parameter group $\Gamma_{\epsilon} = \exp(\epsilon X)$ of transformations. The \emph{prolonged vector field} $X^{(1)}$ on $J^{(1)}$ is the infinitesimal generator of the prolonged 1-parameter group $\Gamma^{(1)}_{\epsilon}$. At each point $(t,y^{(1)}) \in J^{(1)}$ the prolonged vector field is then given by
\begin{equation}
\left. X^{(1)} \right\vert_{(t,y^{(1)})}= \left. \frac{\dd}{\dd\epsilon} \left( \Gamma^{(1)}_{\epsilon} (t,y^{(1)}) \right) \right\vert_{\epsilon=0}.
\end{equation}
\hfill $\square$
\end{definition}

% ---------------------------------------------------------------------

In terms of the prolonged generator $X^{(1)}$, the action of the prolonged 1-parameter group $\Gamma^{(1)}_{\epsilon}$ on jet space $J^{(1)}$ is then given by
\begin{equation}
\Gamma^{(1)}_{\epsilon}(t,y,y') = \left( \hat{t},\hat{y},\hat{y}' \right) = \left( \exp(\epsilon X^{(1)})t , \exp(\epsilon X^{(1)})y , \exp(\epsilon X^{(1)})y' \right),
\end{equation}
and the vector field $X^{(1)}$ can be expressed in components as
\begin{equation}
X^{(1)} = X + \eta^{(1)}_1(t,y,y') \partial_{y'_1} + \ldots + \eta^{(1)}_k(t,y,y') \partial_{y'_k},
\end{equation}
where
\begin{equation}
\eta^{(1)}_i(t,y,y') = \left. \frac{\dd \hat{y}'_i}{\dd \epsilon} \right\vert_{\epsilon=0}, \quad  \quad i=1,\ldots,k.
\end{equation}
Using the total derivative $D_t$, the component functions $\eta^{(1)}_i$ can be expressed directly in terms of the components of $X$ as
\begin{equation}
\label{eqn:prolonged_tangents}
\eta^{(1)}_i = D_t \eta_i - y'_i D_t \xi.
\end{equation}
The infinitesimal description in terms of $X^{(1)}$ greatly facilitates the description and analysis of the prolonged action of a (connected) transformation group $G$. Similarly, an immediate consequence of the general result given in Equation~\eqref{eqn:infinitesimal_invariance} is that a function $I : J^{(1)} \to \mathbb{R}$ is a first order differential invariant for $G$ if and only if
\begin{equation}
\label{eqn:prolonged_infinitesimal_invariance}
X^{(1)}(I) = 0, \quad \quad \forall \, X \in \mathfrak{g}_E.
\end{equation}

% ---------------------------------------------------------------------

\subsection{Symmetries of differential equations}

Equipped with the description of a system of ODEs in jet space $J^{(1)}$ and the induced action of point transformations on $J^{(1)}$ through prolongations, we are now in a position to give a rigorous definition of symmetries of ODEs and provide corresponding infinitesimal formulations. The infinitesimal description makes the symmetries ameneable to analysis by harnessing the fundamental properties of Lie groups, as discussed in the previous sections.   

% ---------------------------------------------------------------------

\begin{definition}
The point transformation $\Gamma : E \to E$ is a \emph{symmetry}\footnote{This class of symmetries is sometimes referred to as point symmetries, to indicate that $\Gamma$ is a point transformation. Since we consider exclusively point transformations, however, we will drop the qualifier \emph{point} and simply use \emph{symmetry}.} of the system $\Delta$ if every solution $y = f(t)$ is mapped to another solution $\hat{y} = (\Gamma f)(\hat{t})$, that is if
\begin{equation}
\label{eqn:symmetry_condition_function}
\Delta(t,f^{(1)}(t)) = 0 \quad \Rightarrow \quad \Delta(\hat{t},(\Gamma f)^{(1)}(\hat{t}))=0,
\end{equation}
or equivalently if
\begin{equation}
\label{eqn:symmetry_condition_graph}
\gamma^{(1)}_f \subset S_{\Delta} \quad \Rightarrow \quad \gamma^{(1)}_{\Gamma f} \subset S_{\Delta}.
\end{equation}
\hfill $\square$
\end{definition}

% ---------------------------------------------------------------------

From the definition above, and that of the prolonged transformation $\Gamma^{(1)}$ in Equation~\eqref{eqn:prolonged_transformation}, it follows immediately that if the prolongation of a point transformation $\Gamma$ preserves the subvariety $S_{\Delta}$, i.e.~$\Gamma^{(1)}(S_{\Delta}) \subset S_{\Delta}$, then $\Gamma$ is a symmetry of the system $\Delta$.

% ---------------------------------------------------------------------

\begin{definition}
The transformation group $G$ acting on $E$ is a \emph{symmetry group} of the system $\Delta$ if $\Gamma_g$ is a symmetry of $\Delta$ for every $g \in G$.
\hfill $\square$
\end{definition}

% ---------------------------------------------------------------------

An important special case of point symmetries and symmetry groups are ones which leave every solution of $\Delta$ invariant, i.e.~mapped to itself under the action of the symmetry. We refer such transformations and groups as \textit{trivial} since they act trivially on the space of solutions\footnote{Note, however, that the action on $S_{\Delta}$ is generally non-trivial.} to $\Delta$.

% ---------------------------------------------------------------------

\begin{figure}[htbp!]
\begin{center}
    \includegraphics[width=0.8\textwidth]{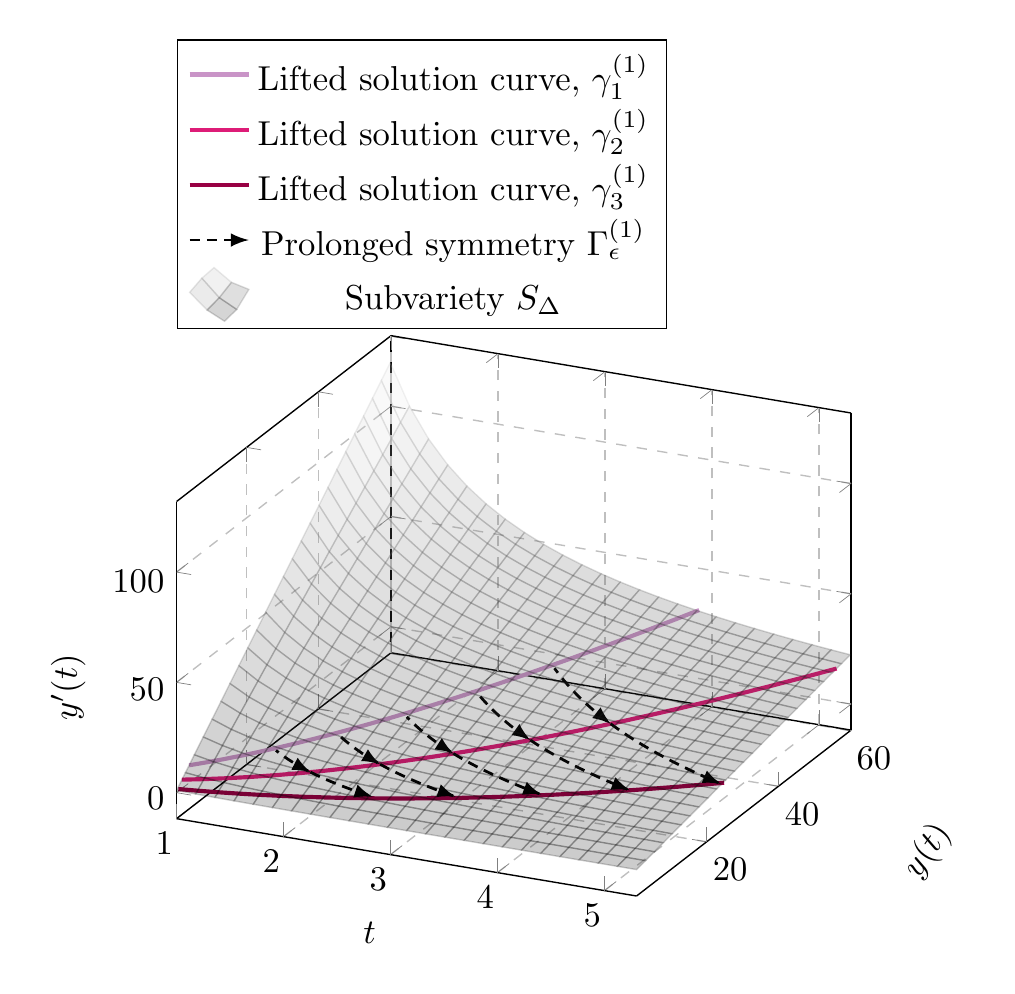}
    \caption{An illustration of the action of the prolonged symmetry transformation $\Gamma^{(1)}_{\epsilon}$ in Equation~(\ref{eq:action_prolonged_symmetry_example}) on the subvariety $S_{\Delta}$ corresponding to the ODE in Equation~(\ref{eqn:single_ode_example}). The lifted solution curves in Figure~\ref{fig:prolongations_example} are mapped to each other by the prolonged symmetry transformation $\Gamma^{(1)}_{\epsilon}$ with parameter $\epsilon = \ln(2)/3$ according to $\gamma^{(1)}_2 = \Gamma_{\epsilon}^{(1)}\gamma^{(1)}_1$ and $\gamma^{(1)}_3 = \Gamma_{\epsilon}^{(1)}\gamma^{(1)}_2 = \left(\Gamma_{\epsilon}^{(1)}\right)^2\gamma^{(1)}_1$.}
\label{fig:prolongation_symmetries_example}
\end{center}
\end{figure}

\begin{example}
Consider again the ODE given in Equation~\eqref{eqn:single_ode_example} and the point transformation generated by the vector field $X = t\partial_t - y\partial_y$, acting on $E$ according to $\Gamma_{\epsilon}(t,y) = (e^{\epsilon}t,e^{-\epsilon}y)$. The components of $X$ are $\xi(t,y)=t$ and $\eta(t,y) = -y$, and computing
\begin{equation}
\eta^{(1)}(t,y,y') = D_t\eta(t,y) - y'D_t\xi(t,y) = -2y', 
\end{equation}
we obtain the prolongation $X^{(1)}$ as
\begin{equation}
X^{(1)} =  t\partial_t - y\partial_y - 2y'\partial_{y'},
\end{equation}
and the corresponding prolonged action on the jet space $J^{(1)}$ as
\begin{equation}
\label{eq:action_prolonged_symmetry_example}
\Gamma^{(1)}_{\epsilon}(t,y,y') = (e^{\epsilon}t,e^{-\epsilon}y,e^{-2\epsilon}y').
\end{equation}
The action of $\Gamma^{(1)}$ on the prolonged graph $\{(t,C_1t^2,2C_1t)\}$ of a solution $y=C_1t^2$ is
\begin{equation}
\Gamma^{(1)}(t,C_1t^2,2C_1t) = (e^{\epsilon}t,e^{-\epsilon}C_1t^2,e^{-2\epsilon}2C_1t) = \left(\hat{t},(e^{-3\epsilon}C_1)\hat{t}^2,2(e^{-3\epsilon}C_1)\hat{t}\right),
\end{equation}
meaning that the transformed function $\Gamma_{\epsilon} f$ according to (\ref{eqn:transformation_action_prolonged_graph}) is given by
\begin{equation}
(\Gamma_{\epsilon} f)(t) = C_{\epsilon} t^2,
\end{equation}
with $C_{\epsilon} = e^{-3\epsilon}C_1$. Clearly, $\Gamma_{\epsilon}f$ is also a solution, meaning that $\Gamma_{\epsilon}$ is indeed a symmetry of the model given in Equation~\eqref{eqn:single_ode_example}. The transformation of solutions, and the corresponding invariance of the subvariety $S_{\Delta}$, is illustrated in Figure~\ref{fig:prolongation_symmetries_example}. 
\hfill $\square$
\end{example}

% ---------------------------------------------------------------------------------

% ---------------------------------------------------------------------

The study of symmetries of ODEs can be approached from different directions, depending on the intended application. We begin by considering the problem of finding the full symmetry group of a given system $\Delta$ of ODEs. The infinitesimal equivalent of the symmetry condition given in Equation~\eqref{eqn:symmetry_condition_function} is provided by the following theorem, whose proof uses properties of group actions on manifolds beyond the scope of the present review.

% ---------------------------------------------------------------------

\begin{theorem}[{\cite[Thm. 6.5]{olver2008equivalence}}]
\label{thm:symmetry_condition_infinitesimal}

Let $G$ be a connected transformation group acting on $E$ and $\Delta$ a regular system of ODEs. Then $G$ is a symmetry group of $\Delta$ if and only if
\begin{equation}
\label{eqn:symmetry_condition_infinitesimal}
\left. X^{(1)}(\Delta) \right\vert_{\Delta=0} = 0, \quad \quad \forall \, X \in \mathfrak{g}_E.
\end{equation}
\hfill $\square$
\end{theorem}  

% ---------------------------------------------------------------------

In components, the condition $X^{(1)}(\Delta) = 0$ amounts to the \emph{determining equations}
\begin{equation}
X^{(1)}(\Delta_i) = 0, \quad \quad i=1,\ldots,k,
\label{eqn:det_eq}
\end{equation}
of the symmetry group. Given a system $\Delta$, solving these equations for the components of $X$, under the assumption $\Delta=0$, amounts to finding the generators $X$ of the full symmetry group $G$ of the system. The group $G$ itself can then be recovered through exponentiation of the Lie algebra $\mathfrak{g}_E$.

Any system of first order ODEs possesses a 1-parameter group of trivial symmetries generated by the vector field defined by the reaction terms in Equation~\eqref{eqn:ode_system}
\begin{equation}
\label{eq:trivial_gen_ode}
    X = \kappa(t,y) \left[\partial_t + \omega_1(t,y)\partial_{y_1} + \ldots + \omega_k(t,y)\partial_{y_k}\right], 
\end{equation}
where $\kappa(t,y)$ is an arbitrary function~\cite{bluman1989symmetries}. In particular, with $\kappa=1$, the vector field generates translations along the solution curves implying that the corresponding symmetries are manifestly trivial. Notwithstanding this triviality, the vector field $X$ (with $\kappa=1$) in Equation~\eqref{eq:trivial_gen_ode} plays an important role in applications through its interpretation as a Hamiltonian vector field for the system $\Delta$ which we will return to below. 

The converse of the problem of finding symmetries to a given system, is to determine the most general system $\Delta$ which admits a given symmetry group $G$. The solution to this problem requires the extension of the general results for differential invariants to the subvariety $S_{\Delta}$ defined by the system, provided by the following theorem.

% ---------------------------------------------------------------------

\begin{theorem}[{\cite[Thm. 6.25]{olver2008equivalence}}]
\label{thm:symmetry_condition_invariants}

Let $G$ be a transformation group whose prolongation $G^{(1)}$ acts regularly with a complete set of functionally independent invariants $I_1,\ldots,I_{\mu_1}$ on an open subset $\Omega^{(1)} \subset J^{(1)}$. Then $G$ is a symmetry group of a system $\Delta$ of ODEs if and only if
\begin{equation}
\label{eqn:symmetry_condition_invariants}
\Delta(t,y^{(1)}) = H(I_1(t,y^{(1)}),\ldots,I_{\mu_1}(t,y^{(1)})) = 0, \quad \quad \forall \, (t,y^{(1)}) \in \Omega^{(1)},
\end{equation} 
for some function $H : J^{(1)} \to \mathbb{R}^k$.
\hfill $\square$
\end{theorem}  

% ---------------------------------------------------------------------

Since differential invariants of $G$ can be found by infinitesimally solving the system in Equation~\eqref{eqn:prolonged_infinitesimal_invariance}, a complete characterisation of systems of ODEs admitting the symmetry group $G$ is obtained from the second equality in Equation~\eqref{eqn:symmetry_condition_invariants} by considering the components $H_i : J^{(1)} \to \mathbb{R}$, with $i=1,\ldots,k$, to be arbitrary functions.

% The solution algorithm
\section{Symbolic symmetry calculations: an algorithm for finding symmetries of first order ODEs}
\label{sec:algorithm}
Since finding the symmetries of any system of differential equations entails solving a high-dimensional system of partial differential equations (PDEs), it is highly desirable to employ computer algebra to perform these calculations. To this end, we present an algorithm for finding the symmetries of a particular class of first order ODEs as in Equation~\eqref{eqn:ode_system} and provide an open-source implementation of the algorithm. We restrict our attention to systems where the reaction terms $\omega_i(t,y)$ for $i=1,\ldots,k$ on the right-hand sides of the ODEs are rational functions of the variable $t$ as well as the states $y_i$, i.e.~of both the independent and the dependent variables. In the context of mathematical biology, this restriction is well-motivated, as the reaction terms of numerous models are based on, for example, logistic growth, mass action kinetics or Michaelis--Menten kinetics, all of which are described by rational reaction terms.

In order to find the infinitesimal generators $X$ in Equation~\eqref{eqn:generating_vector_field} of symmetries we must solve the determining equations in Equation~\eqref{eqn:det_eq}. In practice, this is accomplished by first using the linearity of the prolonged infinitesimal generator $X^{(1)}$ to express the determining equations
\begin{equation}
X^{(1)} \left( y'_i - \omega_i(t,y) \right) = 0 , \quad i=1,\ldots,k,
\end{equation}
in terms of the generator $X = \xi(t,y)\partial_t + \eta_1(t,y)\partial_{y_1} + \cdots + \eta_k(t,y)\partial_{y_k}$ and its components as
\begin{equation}
\label{eq:det_eq_practice}
D_t\eta_i-\omega_i D_t\xi=X\left(\omega_i(t,y)\right),\quad i=1,\ldots,k \,.
\end{equation}
Here, we have used the expression in Equation~\eqref{eqn:prolonged_tangents} for the components of the prolonged generator and the fact that $X^{(1)}(\omega_i(t,y)) = X(\omega_i(t,y))$ for all $i=1,\ldots,k$. Note that this is a system of $k$ non-linear PDEs in $k+1$ variables, and thus the difficulty of finding the symmetries scales linearly with the dimensionality of the ODE system of interest.

In order to construct an algorithm for solving the determining equations with rational reaction terms $\omega_i$, we will use a set of ans\"atze for the components $\xi$ and $\eta_i$ of the infinitesimal generator $X$. Specifically, we restrict the components of the infinitesimal generator to be \textit{polynomial} in the states which results in a linear system of equations in the coefficients $c_{ij}(t)$ of the monomials appearing in the ans\"atze, where the index $i$ corresponds to the number of components and the index $j$ to the number of distinct monomials. In general, the number $n$ of unknown coefficients can be calculated by the degree $d$ of the polynomials in the ans\"atze and the number of states $k$ according to  
\begin{equation}
n=(k+1)\binom{k+d}{d}.
\label{eq:number_of_equations}
\end{equation}

% --------------------------------------------------------------------

\begin{example}
We exemplify the notation for the ansatz and the resulting determining equations by considering a two component system of ODEs,~i.e. where $k=2$, with an ansatz of degree $d=1$. The infinitesimal generator $X$ is then 
\begin{equation}
X=\xi(t,y_1,y_2)\partial_{t}+\eta_1(t,y_1,y_2)\partial_{y_1}+\eta_1(t,y_1,y_2)\partial_{y_2},
\end{equation}
with component ans\"atze of the form
\begin{align}
\xi(t,y_1,y_2)&=c_{00}(t)+c_{01}(t)y_1+c_{02}(t)y_2,\\
\eta_1(t,y_1,y_2)&=c_{10}(t)+c_{11}(t)y_1+c_{12}(t)y_2,\\
\eta_2(t,y_1,y_2)&=c_{20}(t)+c_{21}(t)y_1+c_{22}(t)y_2.
\end{align}
The determining equations, expressed in terms of the nine unknown coefficient functions $c_{ij}(t)$, are then given by
\begin{equation}
\begin{array}{rcl}
X\left(\omega_i(t,y_1,y_2)\right) & = &
\left(c'_{i0} + c'_{i1}y_1 + c'_{i2}y_2 + c_{i1}\omega_1(t,y_1,y_2) + c_{i2}\omega_2(t,y_1,y_2) \right) \\ & & - \omega_i(t,y_1,y_2) \left( c'_{00} + c'_{01}y_1 + c'_{02}y_2 + c_{01}\omega_1(t,y_1,y_2) + c_{02}\omega_2(t,y_1,y_2) \right)
\end{array}
\end{equation}
for $i=1,2$, where the left-hand sides are linear in the coefficients and their derivatives. The number of independent equations is determined by the form of the reaction terms $\omega_1(t,y_1,y_2)$ and $\omega_2(t,y_1,y_2)$.
\hfill $\square$
\end{example}

% --------------------------------------------------------------------

In general, inserting the polynomial ans\"atze into the determining equations given in Equation~\eqref{eq:det_eq_practice} for a system with rational reaction terms yields a linear system of ODEs for the unknown coefficients $c_{ij}(t)$ in the tangential ans\"atze. This system can be formulated as a matrix system 
\begin{equation}
A(t)\dv{\mathbf{c}}{t}=B(t)\mathbf{c}(t),
\label{eq:det_eq_matrix_form}
\end{equation}
where $\mathbf{c}(t)$ is the $n$-dimensional continuous vector-valued function consisting of all the unknown coefficients in the tangential ans\"atze that we want to solve for. The time-dependent matrices $A$, $B$ have dimensions $m\times n$ where the number of columns $n$ is given by the number of coefficients in Equation~\eqref{eq:number_of_equations}, and the number of equations $m$ is typically much larger than the number of unknowns, i.e. $m\gg n$. Specifically, the number of equations $m$ depends on the degree of the polynomials in the reaction terms as well as the degree $d$ of the polynomials in the tangential ans\"atze.

In other words, the resulting system is an \textit{overdetermined} linear system of first order ODEs in the coefficients $c_{ij}(t)$ which can be solved by reducing the system in Equation~\eqref{eq:det_eq_matrix_form} to an inhomogeneous quadratic matrix system and an auxiliary set of algebraic equations
\begin{equation}
\dv{\mathbf{c}}{t}=B_{\text{diff}}\,\mathbf{c}(t) + \mathbf{d}_1(t), \qquad B_{\text{alg}}(t)\mathbf{c}(t) + \mathbf{d}_2(t) = 0,
\label{eq:det_eq_reduced_matrix_form}
\end{equation}
where $B_{\text{diff}}$ is a constant matrix, $\mathbf{d}_1(t)$ and $\mathbf{d}_2(t)$ contain the inhomogeneities remaining after the reduction and $B_{\text{alg}}$ encodes the algebraic equations. The general solution to the quadratic system of differential equations can be obtained using the Jordan decomposition of $B_{\text{diff}}$, and the algebraic constraints are subsequently applied to arrive at the solution to the original system given in Equation~\eqref{eq:det_eq_matrix_form}.

We have developed an open-source implementation of this algorithm, using the symbolic solver \textit{SymPy} \cite{meurer2017sympy} to extract the matrix system given in Equation~\eqref{eq:det_eq_matrix_form}, perform the reduction and solve the resulting differential and algebraic equations. Our implementation, as well as the details of the algorithm, are provided in the public repository associated with this work (see \url{https://github.com/JohannesBorgqvist/symSys_1st_ODEs}). The algorithm will always generate a solution to Equation~\eqref{eq:det_eq_matrix_form} but in general only the trivial solution $\mathbf{c} = \mathbf{0}$, corresponding to the case where no generators of the form given by the polynomial ans\"atze exist. If non-trivial solutions exist, the coefficients $\mathbf{c}(t)$ are substituted back into the tangential ans\"atze in order to produce the resulting infinitesimal generators. Although this algorithm is by no means guaranteed to find generators of the particular form of the ansatz, it is scalable since it amounts to solving the linear system in Equation~\eqref{eq:det_eq_matrix_form}. 

In principle, the algorithm allows us to test numerous degrees in the polynomial ans\"atze and it provides an indispensable tool in the systematic search for symmetries of systems of ODEs with multiple states. In practice, however, symbolic calculations are notoriously slow and so to explore the symmetries of models consisting of large systems of ODEs with high order ans\"atze the algorithm requires efficient implementation and the use of high performance symbolic calculations. We emphasise that our implementation is by no means optimised for performance, however we have successfully applied it to calculate the infinitesimal generators of a range of selected models. Here we present two initial examples of models of biological relevance where the infinitesimal generators were successfully obtained.

% --------------------------------------------------------------------------------------

\subsection{Example: Hydon's model}

As a first example of the application of the algorithm described above, we consider the following nonlinear system of ODEs which will be referred to as \textit{Hydon's model} \cite{hydon2000symmetry}: 
\begin{equation}
\label{eq:hydon}
\begin{split}
\dv{y_1}{t} &= \omega_1(t,y_1,y_2) = \frac{t y_{1} + y_{2}^{2}}{y_{1} y_{2}- t^{2} },\\
\dv{y_2}{t} &= \omega_2(t,y_1,y_2) = \frac{t y_{2} + y_{1}^{2}}{y_{1} y_{2}- t^{2} }.
\end{split}
\end{equation}
The algorithm applied with an ansatz of degree $d=2$ produces the two generators 
\begin{align}
X_1 &= t\partial_t + y_1\partial_{y_1} + y_2\partial_{y_2}, \label{eq:hydon_gen_1}\\
X_2 &= \kappa(t)\left[(y_1 y_2 - t^2)\partial_t + (t y_1 + y_2^2)\partial_{y_1} + (t y_2 + y_1^2 )\partial_{y_2}\right]. \label{eq:hydon_gen_2}
\end{align}
The first of the two generators, i.e.~$X_1$ in Equation~\eqref{eq:hydon_gen_1}, is non-trivial and known to be the only existing generator linear in both states $y_1$ and $y_2$. The second generator, i.e.~$X_2$ in Equation~\eqref{eq:hydon_gen_2}, is parallel to the vector field defined by the reaction terms in the system of ODEs, and hence acts trivially on the space of solutions. Also, we note that $X_2$ corresponds to a family of generators, due to the overall scaling by an arbitrary function $\kappa(t)$.

\begin{figure}[htbp!]
\begin{center}
\includegraphics[width=0.8\textwidth]{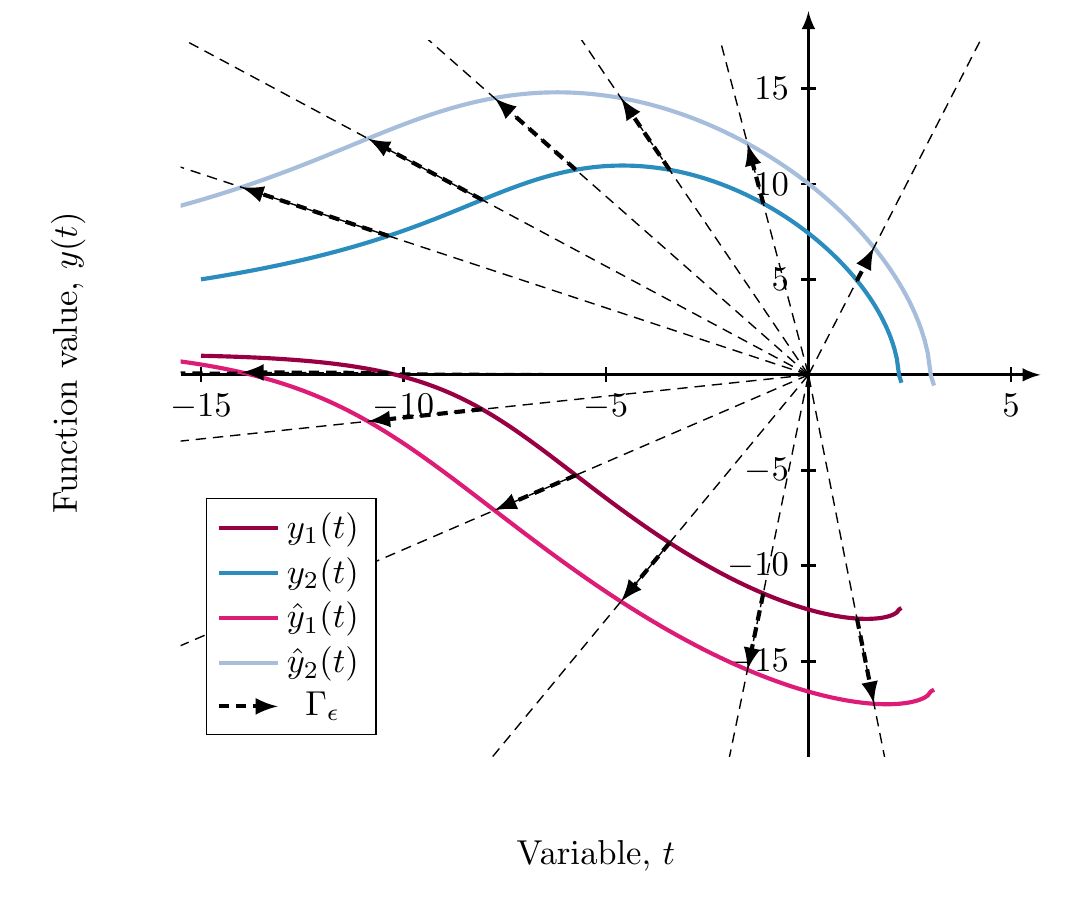}
\caption{An illustration of the action of the three-dimensional scaling symmetry $\Gamma_{\epsilon}$ in Equation~\eqref{eq:scaling_symmetry} on the total $(t,y_1,y_2)$-space of solutions to the two component system of ODEs in Equation~\eqref{eq:hydon}. The two solutions $(y_1(t),y_2(t))$ and $(\hat{y}_1(t),\hat{y}_2(t))$ are related through the action of $\Gamma_{\epsilon}$ with parameter $\epsilon=0.3$. Also, the scaling symmetry $\Gamma_{\epsilon}$ is generated by $X_1$ in Equation~\eqref{eq:hydon_gen_1} and the orbits of $\Gamma_{\epsilon}$ given by the vector field $X_1$ are illustrated by the thin dashed lines.}
\label{fig:hydons_model}
\end{center}
\end{figure}

In order to visualise the action of the non-trivial generator in Equation~\eqref{eq:hydon_gen_1}, we exponentiate $X_1$ to obtain the corresponding symmetry transformation given by
\begin{equation}
    \Gamma_{\epsilon}(t,y_1,y_2) = \left(e^{\epsilon}t,e^{\epsilon}y_1,e^{\epsilon}y_2\right).
    \label{eq:scaling_symmetry}
\end{equation}
The transformation in Equation~\eqref{eq:scaling_symmetry} corresponds to a simultaneous scaling of both independent and dependent variables, as illustrated in Figure~\ref{fig:hydons_model}, and is therefore referred to as a \textit{scaling symmetry} of the system of ODEs given in Equation~\eqref{eq:hydon}.

% ---------------------------------------------------------------------

% ----------------------------------------------------------------

\subsection{Example: A linear model}

Another example is provided by the special case of a linear two-state model given by
\begin{equation}
\label{eq:lin}
\begin{split}
\dv{y_1}{t} &= \omega_1(t,y_1,y_2) = y_1+y_2,\\
\dv{y_2}{t} &= \omega_2(t,y_1,y_2) = y_1+y_2.
\end{split}
\end{equation}
Biologically, this system of ODEs describes synergistic growth of, for example, two populations of cells denoted by $y_1(t)$ and $y_2(t)$. While this specific model does not describe interacting populations dynamics realistically, linear models have numerous applications in general and in particular they occur in the context of modelling complex dynamics in systems biology.

Using a set of tangential ans\"atze of degree $d=1$, our implementation of the algorithm finds nine functionally independent infinitesimal generators of symmetries of the ODEs in Equation~\eqref{eq:lin}, which can be cast on the form
\begin{align}
X_{1} &= (-y_1 + y_2)\partial_t,\\
X_{2} &= (y_1 + y_2)e^{-2t}\partial_t,\\
X_{3} &= -\partial_{y_1} + \partial_{y_2},\\
X_{4} &= e^{2t}\partial_{y_1} + e^{2t}\partial_{y_2},\\
X_{5} &= y_1\partial_{y_1} + y_2\partial_{y_2},\label{eq:lin_1}\\
X_{6} &= y_2\partial_{y_1} + y_1\partial_{y_2},\label{eq:lin_2}\\
X_{7} &= e^{2t}(y_1-y_2)\partial_{y_1} + e^{2t}(y_1-y_2)\partial_{y_2},\\
X_{8} &= e^{-2t}(y_1+y_2)\partial_{y_1} - e^{-2t}(y_1+y_2)\partial_{y_2},\\
X_{9} &= \kappa(t)\left[\partial_t + (y_1+y_2)\partial_{y_1} + (y_1+y_2)\partial_{y_2}\right],
\end{align}
using suitable linear combinations, where again $\kappa(t)$ appearing in the trivial generator $X_9$ is an arbitrary function.

In fact, the generator $X_5$ is a symmetry generator for the general linear model, meaning that it is common to all linear two component system of ODEs. Furthermore, from the form of the reaction terms in Equation~\eqref{eq:lin}, which are autonomous (i.e.~have no explicit dependence on the independent variable $t$), we observe that the vector field $\partial_t$ generates a manifest \textit{translation symmetry}\footnote{A large class of models in mathematical biology consist of autonomous ODEs making the time translation generated by $\partial_t$ a frequently occurring symmetry.}
\begin{equation}
    \Gamma_\epsilon(t,y_1(t),y_2(t)) = (t+\epsilon,y_1(t),y_2(t)).
    \label{eq:translation_symmetry}
\end{equation}
Indeed, by taking $\kappa(t)=1$ this generator can be obtained as the linear combination
\begin{equation}
    \partial_t = X_9 - X_5 - X_6.
\end{equation}
We will consider both generators $X_5$ and $\partial_t$ in greater detail when we derive models starting from the symmetries. Before that, we will proceed to discuss the interpretation of symmetries of some well-known biological models.

% Symmetries found using the algorithm
\section{Understanding biological models: inferring biophysical properties from symmetries}
\label{sec:bioexamples}
Given our algorithm for finding symmetries, we will now analyse two well-known models in mathematical biology, namely the SIR model and the Lotka--Volterra model. More specifically, we present the symmetries that were calculated using our algorithm, and then we interpret their meaning in terms of biological properties of the underlying systems. In biological applications, we want to be able to distinguish between the time and the state space variables for the solutions of the model which, in the language of symmetries, means that we are interested in maintaining the fibration structure. In order for symmetries to preserve this structure of fibrations over time, we restrict our attention to generators which are \emph{projective}, meaning that the components of these infinitesimal generators $X$ in the $t$-direction satisfy $\xi=\xi(t)$. To elucidate biologically relevant properties of the two models of interest, we investigate the \emph{invariants}, see Equation~\eqref{eqn:prolonged_infinitesimal_invariance}, of their projective generators which correspond to conserved quantities of the models. By interpreting these conserved quantities biologically, we show how well-known properties of the SIR and Lotka--Volterra models emerge from their respective symmetries.

% ---------------------------------------------------------------------

\subsection{The SIR model: mass conservation and autonomy}

The SIR model consists of the following three-state system of first order ODEs
\begin{equation}
\begin{split}
\frac{\dd S}{\dd t} &= \omega_S(t,S,I,R) = -ISr,\\
\frac{\dd I}{\dd t} &= \omega_I(t,S,I,R) = ISr - Ia,\\
\frac{\dd R}{\dd t} &= \omega_R(t,S,I,R) = Ia,\\        
\end{split}
\label{eq:SIR}
\end{equation}
which describe the dynamics of the spread of an infectious disease in a population subdivided into susceptible, $S(t)$, infected, $I(t)$, and recovered, $R(t)$, individuals. The projective infinitesimal generators that were calculated using our algorithm with a set of tangential ans\"atze of degree $d=2$ are
\begin{align}
X_1 &= \partial_t, \label{eq:translation}\\
X_{2} &= \partial_R,
\label{eq:SIR_R_transl}\\
X_{3} &= (S+I+R)\partial_R,\label{eq:SIR_strange_generator}\\
X_{4} &= \kappa (t)\left[\partial_t - rSI\partial_S + \left(rSI - aI\right)\partial_I + aI\partial_R\right].
\label{eq:SIR_trivial}
\end{align}
Since these generators are projective (they do not mix the states with the time), we consider them to be biologically relevant. Beginning with the infinitesimal generator $X_1 = \partial_t$, we note that this is the well-known translation generator. It is manifest in the SIR model in Equation~\eqref{eq:SIR} as the system is autonomous, meaning that its reaction terms have no explicit time-dependence. The prolonged generator is $X_1^{(1)} = X_1 = \partial_t$ and the corresponding prolonged group $G^{(1)}$ acts on $J^{(1)}$ with orbits of dimension $s_1 = 1 = \dim{G}$. Consequently, there are $\mu_1 = \dim{J^{(1)}} - s_1 = 7 - 1 = 6$ first order differential invariants according to Equation~\eqref{eqn:functionally_independent_differential_invariants}, and the characteristic equations corresponding to the invariance condition in  Equation~\eqref{eqn:differential_invariant} are
\begin{equation}
\frac{\dd t}{\dd s} = 1, \quad \quad \frac{\dd S}{\dd s} = \frac{\dd I}{\dd s} = \frac{\dd R}{\dd s} = \frac{\dd S'}{\dd s} = \frac{\dd I'}{\dd s} = \frac{\dd R'}{\dd s} = 0.
\end{equation}
It is straightforward to show that the invariants of $X_1$ are all states and their derivatives
\begin{equation}
I_1 = S, \qquad I_2 = I, \qquad I_3 = R, \qquad I_4 = S', \qquad I_5=I', \qquad I_6 = R'.
\end{equation}
Thus, symmetry under $X_1$ amounts to the invariance of the states as well as invariance of the dynamics of the model under time translations. Or, more colloquially formulated, the biological mechanisms governing the system are the same no matter when an experiment is conducted. In other words, shifting a solution in time produces another solution corresponding to a different set of initial conditions.

Next, we consider the generator $X_2 = \partial_R$ in Equation~\eqref{eq:SIR_R_transl} that arises because all reaction terms in Equation~\eqref{eq:SIR} are independent of the state $R$. Once again, $\mu_1=6$ and the computation of the invariants is straightforward, yielding
\begin{equation}
\label{eq:SIR_R_transl_invariants}
I_1 = t, \qquad I_2 = S, \qquad I_3 = I, \qquad I_4 = S', \qquad I_5=I', \qquad I_6 = R'.
\end{equation}
Symmetry under $X_2$ amounts to the fact that a change in the state $R$ leaves the remaining states $S$ and $I$ unchanged. More importantly, a change in the state $R$ does not affect the dynamics of the model. In the context of the SIR model, this corresponds to the population of recovered individuals having no influence on the spread of the disease since there is no feedback on the susceptible or infected states. That is, once an individual is recovered they will remain recovered at all later times. When restricted to the space of solutions of Equation~\eqref{eq:SIR}, the invariants of the generator $X_3$ in Equation~\eqref{eq:SIR_strange_generator} constitute a subset of the ones in Equation~\eqref{eq:SIR_R_transl_invariants} and we will therefore not consider it further here. 

The final biologically relevant generator is the generator $X_4$ in Equation~\eqref{eq:SIR_trivial}. By fixing the arbitrary function to $\kappa(t) = 1$, this generator can be written as follows
\begin{equation}
X_4 = \partial_t - rSI\partial_S + (rSI-aI)\partial_I + aI\partial_R.
\end{equation}
This infinitesimal generator is \emph{trivial} in the sense that it generates translations along solution curves. More specifically, since the vector field defining the right-hand sides of the SIR model in Equation~\eqref{eq:SIR} is given by
\begin{equation}
X_4 = \partial_t + \omega_S\partial_S + \omega_I\partial_I + \omega_R\partial_R,
\end{equation}
it is clear that the vector field of the trivial infinitesimal generator $X_4$ in Equation~\eqref{eq:SIR_trivial} is \emph{parallel} to the vector field of the model itself. This implies that the symmetry transformation generated by $X_4$ maps points \emph{along the same solution curve}. However, even though the action of $X_4$ on the solution space is trivial in the sense of symmetries, it contains biologically relevant information. In particular, the invariants of the trivial vector field correspond to quantities of the system conserved during its time evolution. In this case, it can be shown that the total population size, $N = S + I + R$, is an invariant of $X_4$, implying that it is a constant for each solution of the model. This is the well-known mass-conservation property of the SIR model. Furthermore, all trivial generators on the form $X_4$ in Equation~\eqref{eq:SIR_trivial} obtained by multiplication by an arbitrary function $\kappa(t)$ act trivially on the space of solutions and in fact they comprise the so-called null space of the Lie algebra.

This example of the SIR model illustrates the different roles that symmetries can play in furthering understanding of the dynamics of a model. The infinitesimal generators given by the time translation generator $X_1$ in Equation~\eqref{eq:translation}, the $R$-translation generator $X_3$ in Equation~\eqref{eq:SIR_R_transl} and the trivial generator $X_4$ in~\eqref{eq:SIR_trivial} can be understood in terms of familiar properties of the system in Equation~\eqref{eq:SIR}. Interestingly, non-trivial symmetries provide information on the dynamics by relating different solutions to each other, while the trivial symmetry generator, referred to as the Hamiltonian vector field in classical mechanics, is associated with \emph{conservation laws} of the model. 

% ---------------------------------------------------------------------

\subsection{The Lotka--Volterra model: energy conservation}

The Lotka--Volterra model, in its dimensionless form, is described by the following two-state system of first order ODEs
\begin{equation}
\begin{split}
\dv{u}{t} &= \omega_u(t,u,v) = u(1-v),\\
\dv{v}{t} &= \omega_v(t,u,v) = v(u-1).
\end{split}
\label{eq:LV}
\end{equation}
Here, the change of two interacting populations of prey, $u(t)$, and predators, $v(t)$, over time is described. Again, using the proposed algorithm with a set of tangential ans\"atze of degree $d=2$, we calculate the following generators
\begin{align}
X_{1} &= \partial_t,\\
X_{2} &= \kappa(t)\left[\partial_t+u(1-v) \partial_u+av(u-1)\partial_v\right].\label{eq:LV_trivial}
\end{align}
Similarly to the SIR model, the Lotka-Volterra model possesses a manifest time translation symmetry $X_1$ previously presented in Equation~\eqref{eq:translation}. Dimensional considerations show that there are $\mu_1 = \dim{J^{(1)}} - s_1 = 5 - 1 = 4$  first order differential invariants for this two-state model that, in analogy with the SIR model, are given by
\begin{equation}
I_1 = u, \qquad I_2 = v, \qquad I_3 = u', \qquad I_4 = v'.
\end{equation}
The biophysical interpretation is also analogous to that of the SIR model; the absence of explicit time-dependence in the reaction terms entails the invariance of the model dynamics under time translations. 

Moreover, conserved quantities of the Lotka--Volterra model in Equation~\eqref{eq:LV} are obtained by considering the \emph{trivial symmetry} $X_2$ in Equation~\eqref{eq:LV_trivial}. Again, similarly to the analysis of the SIR model, it is clear that this symmetry is trivial because by fixing the arbitrary function to $\kappa(t)=1$ this infinitesimal generator becomes
\begin{equation}
X_2=\partial_t+u(1-v) \partial_u+av(u-1)\partial_v,
\end{equation}
which is a vector field that is parallel to the vector field $X=\partial_t+\omega_u\partial_u+\omega_v\partial_v$ given by the reactions terms in Equation~\eqref{eq:LV}. Furthermore, the corresponding prolonged generator is given by 
\begin{equation}
X_2^{(1)}=X_2+\left[u'(1-v)-uv'\right]\partial_{u'}+\left[av'(u-1)+au'v\right]\partial_{v'}.
\end{equation}
and the invariants of $X_2^{(1)}$ are calculated by solving the equation $X_2^{(1)}(I) = 0$. In particular, this equation can be decomposed into the associated characteristic equations given by
\begin{equation*}
\frac{\dd t}{1} = \frac{\dd u}{u(1-v)} = \frac{\dd v}{av(u-1)} = \frac{\dd u'}{u'(1-v)-uv'} = \frac{\dd v'}{av'(u-1)+au'v} = \dd s.
\end{equation*}
By combining the second and third expressions, we obtain the well-known state space ODE
\begin{equation}
\dv{u}{v}=\dfrac{u(1-v)}{av(u-1)},
\end{equation}
which has the solution 
\begin{equation}
I_1 = a\left(u -\ln u\right) + v - \ln v =  au+v-\ln\left(u^a v\right),
\label{eq:LV_I1}    
\end{equation}
where the invariant $I_1$ appears as an arbitrary integration constant. Consequently, the quantity $I_1$ is conserved along solution trajectories, i.e.~the solutions are level curves of the Hamiltonian function $I_1$. The solutions in Equation~\eqref{eq:LV_I1} represent closed trajectories in state space of constant (generalised) energy~\cite{murray2002}. Thus, symmetries allow us to derive conservation laws through the invariants of the trivial generators. This fact demonstrates that symmetry methods constitute a powerful theoretical tool for analysing the properties of a given model. Yet, an even more promising prospect is to reverse the direction of the analysis, so that instead of analysing a given model by calculating its symmetries, we can start from symmetries in order to derive models. This will allow the physical or biological properties to be built into the very structure of the constructed model, which is the focus of the next section.

% Deriving models using symmetries
\section{Constructing biological models: making biophysical properties manifest using differential invariants}
\label{sec:model_construction}
In contrast to the previous analysis where the symmetries of a given model were calculated, an arguably equally interesting question is what models admit a given group of symmetries? As symmetries correspond to biophysical properties of a system, this implies that the derivation of the most general model compatible with a set of symmetries amounts to encoding those properties in the very structure of the model. In addition, expressing a model of a biological system in such a way as to make its symmetries manifest, i.e.~explicit in the mathematical description of the model, and exhibit the underlying structures, can be very useful in elucidating the underlying biological mechanisms governing the system. This is a hugely promising approach, used with great success in many areas of mathematical physics, as it enables the construction of  robust and interpretable models where the underlying mechanisms of a given biological system are captured through its symmetries. 

For systems of ODEs, the construction of the class of models admitting a group $G$ of symmetries uses the differential invariants, Equation~\eqref{eqn:differential_invariant}, of $G$ and Theorem~\ref{thm:symmetry_condition_invariants} on the invariance of differential equations. To illustrate the application of the theory, and the construction of differential invariants from infinitesimal generators, we will now construct invariant first order ODE models for a number of symmetry groups found when well-known models in mathematical biology were analysed. In the first example, we consider the commonly occurring invariance under time translation $\partial_t$ for a single ODE and for a system of two ODEs. In all of the subsequent examples, we consider models with two states $y_1(t)$ and $y_2(t)$, meaning that we consider ODEs with one independent and two dependent variables ($k=2$).

% --------------------------------------------------------------------

\subsection{Time translation and autonomy}

In the case of one independent and one dependent variable ($k=1$) the components of the generator $X=\partial_t$ are $\xi=1,\,\eta=0$, and the first prolonged component of this infinitesimal generator is given by
\begin{equation}
\eta^{(1)} = D_t\eta-y'D_t\xi = 0,
\end{equation}
where the total derivative $D_t$ is defined in Equation~\eqref{eq:tot_derivative}. Accordingly, the prolonged generator is $X^{(1)}=X=\partial_t$ and the prolonged group $G^{(1)}$ acts regularly on $J^{(1)}$ with orbits of dimension $s_1 = 1 = \mathrm{dim}\,G$. Consequently, according to Equation~\eqref{eqn:functionally_independent_differential_invariants} there are $\mu_1 = \mathrm{dim}\,J^{(1)} - s_1 = 3 - 1 = 2$ first order differential invariants satisfying $X^{(1)}(I) = 0$. The corresponding characteristic equations are
\begin{equation}
\frac{\text{d}t}{1} = \text{d}s, \qquad \frac{\text{d}y}{\text{d}s} = \frac{\text{d}y'}{\text{d}s} = 0,
\end{equation}
implying that a complete set of functionally independent first order invariants of $G^{(1)}$ is given by
\begin{equation}
I_1 = y, \quad I_2 = y'.
\end{equation}
According to Theorem~\ref{thm:symmetry_condition_invariants}, the most general ODE admitting  $G$ as a symmetry group is therefore of the form $H(y,y')=0$ or, solving for the derivative yields
\begin{equation}
y' = F(y),
\label{eq:autonomy_1}
\end{equation}
where $F$ is an arbitrary differentiable function. In other words, the symmetry under translations in the independent variable $t$ is made manifest in the absence of explicit time-dependence in the reaction term $\omega(t,y) = F(y)$.

Extending the example above to a system with $k=2$ ODEs, we have that the components of the infinitesimal generator $X=\partial_t$ are given by $\xi=1$, and $\eta_1=\eta_2=0$. Once again, the prolongations of the tangents are trivial implying that $X^{(1)}=X=\partial_t$ and the orbits of $G^{(1)}$ have dimension $s_1=1$ as before. In this case, there are $\mu_1 = \mathrm{dim}\,J^{(1)} - s_1 = 5 - 1 = 4$ first order differential invariants satisfying $X^{(1)}(I) = 0$ according to Equation~\eqref{eqn:functionally_independent_differential_invariants}. The associated characteristic equations are
\begin{equation}
\frac{\text{d}t}{1} = \text{d}s, \qquad \frac{\text{d}y_1}{\text{d}s} = \frac{\text{d}y_2}{\text{d}s} = \frac{\text{d}y'_1}{\text{d}s} = \frac{\text{d}y'_2}{\text{d}s} = 0,
\end{equation}
and hence a complete set of differential invariants for $G^{(1)}$ are given by
\begin{equation}
I_1 = y_1, \qquad I_2 = y_2, \qquad I_3 = y'_1, \qquad I_4 = y'_2.
\end{equation}
Given these invariants, the most general system invariant under $G$ is given by
\begin{equation}
H_i(y_1,y_2,y'_1,y'_2) = 0, \qquad i=1,2,
\end{equation}
according to Theorem~\ref{thm:symmetry_condition_invariants}, or solving for the derivatives,
\begin{equation}
y'_i = F_i(y_1,y_2), \qquad i=1,2,
\label{eq:autonomy_2}
\end{equation}
for arbitrary functions $F_1$ and $F_2$. These results in Equation~\eqref{eq:autonomy_1} and Equation~\eqref{eq:autonomy_2} generalise in a straightforward manner to additional dependent variables to the equivalence of autonomy and time translation invariance. Consequently, the take-home message is that time translations generated by the infinitesimal generator $X = \partial_t$ are common symmetries of all \emph{autonomous} models. 

% --------------------------------------------------------------------

\subsection{Total space scaling symmetry}

Previously, Hydon's model in Equation~\eqref{eq:hydon} was shown above to possess a single linear infinitesimal generator defined in Equation~\eqref{eq:hydon_gen_1}. This infinitesimal generator is given by 
\begin{equation}
X=t\partial_t + y_1\partial_{y_1} + y_2\partial_{y_2},
\end{equation}
and it generates the scaling symmetry in Equation~\eqref{eq:scaling_symmetry} which is illustrated in Figure~\ref{fig:hydons_model}. Its components are $\xi=t$, $\eta_1=y_1$ and $\eta_2=y_2$, and precisely as in the previous example the corresponding prolonged components are trivial
\begin{equation}
\eta^{(1)}_i = D_t\eta_i-y'_iD_t\xi = 0, \qquad i= 1,2.
\end{equation}
Again, this means that the prolonged infinitesimal generator of the scaling symmetry satisfies $X^{(1)}=X$ or, alternatively, that the prolonged infinitesimal generator of the scaling symmetry coincides with that of $G$. Moreover, the 1-parameter symmetry group $G^{(1)}$ generated by $X^{(1)}$ acts on $J^{(1)}$ through simultaneous rescalings in the total space $E$ according to
\begin{equation}
\Gamma^{(1)}_{\epsilon}(t,y_1,y_2,y'_1,y'_2) = (e^{\epsilon}t,e^{\epsilon}y_1,e^{\epsilon}y_2,y'_1,y'_2).
\end{equation}
The orbits of this action are the individual points in the fibre at the origin of $E$ and the rays emanating from them. Consequently, the generic orbit dimension is $s_1=1=\mathrm{dim}\,G$ and the number of functionally independent first order differential invariants are $\mu_1 = \mathrm{dim}\,J^{(1)} - s_1 = 5 - 1 = 4$ according to Equation~\eqref{eqn:functionally_independent_differential_invariants}. The first order differential invariants of $X^{(1)}$ are obtained by solving the invariance condition $X^{(1)}(I) = 0$ according to Equation~\eqref{eqn:differential_invariant} or, equivalently, finding first integrals of the characteristic system
\begin{equation}
\frac{\text{d}t}{t} = \frac{\text{d}y_1}{y_1} = \frac{\text{d}y_2}{y_2} = \text{d}s, \qquad \frac{\text{d}y'_1}{\text{d}s} = \frac{\text{d}y'_2}{\text{d}s} = 0.
\end{equation}
The resulting complete set of first order differential invariants can be taken as
\begin{equation}
I_1 = \frac{y_1}{t}, \qquad I_2 = \frac{y_2}{t}, \qquad I_3 = y'_1, \qquad I_4 = y'_2.
\label{eq:hydon_invariant}
\end{equation}
According to Theorem~\ref{thm:symmetry_condition_invariants}, the most general first order system invariant under $G$ is given by
\begin{equation}
H_i\left( \frac{y_1}{t} , \frac{y_2}{t} , y'_1 , y'_2 \right) = 0, \qquad i=1,2.
\end{equation}
Once again, we can solve for the derivatives to obtain
\begin{equation}
y'_i = F_i \left( \frac{y_1}{t} , \frac{y_2}{t} \right),\qquad i=1,2,
\label{eq:general_scaling_ODE}
\end{equation}
as the class of invariant models where, as before, $F_1$ and $F_2$ are arbitrary functions.

Specifically, we recover Hydon's model, in Equation~\eqref{eq:hydon}, from the general system of ODEs in Equation~\eqref{eq:general_scaling_ODE} by re-writing its reaction terms in terms of the invariants in Equation~\eqref{eq:hydon_invariant}
\begin{equation}
\omega_1 = \frac{ty_1 + y_2^2}{y_1y_2-t^2} = \frac{\frac{y_1}{t}+\left(\frac{y_2}{t}\right)^2}{\frac{y_1}{t}\frac{y_2}{t}-1}, \quad \omega_2 = \frac{ty_2 + y_1^2}{y_1y_2-t^2} = \frac{\frac{y_2}{t}+\left(\frac{y_1}{t}\right)^2}{\frac{y_1}{t}\frac{y_2}{t}-1},
\end{equation}
corresponding to the choice
\begin{equation}
F_1(x_1,x_2) = \frac{x_1+x_2^2}{x_1x_2 - 1}, \quad F_2(x_1,x_2) = \frac{x_1^2+x_2}{x_1x_2 - 1},
\end{equation}
of the arbitrary functions in Equation~\eqref{eq:general_scaling_ODE}.

In addition to verifying the symmetry of Hydon's model, Equation~\eqref{eq:hydon}, under $G$, this example illustrates that the underlying structure of the model is made more explicit by constructing the model from its symmetries. Thus, by making the symmetries of a model manifest by means of expressing its reaction terms in terms of its differential invariants, we can simultaneously highlight the symmetry of the biological mechanism modelled. Although, Hydon's model in Equation~\eqref{eq:hydon} is not a biological model, we will now repeat this analysis to derive models using the invariants of more biologically motivated symmetries.

% --------------------------------------------------------------

\subsection{State space symmetries}

In models containing numerous dependent variables, symmetries acting non-trivially only on state space $U$ are of particular interest. In many situations, the governing biological mechanisms are not expected to change over time and symmetries restricted to state space can provide information about the system that is isolated from the time-dependence of particular solutions. To exemplify, we consider the most general class of two state models associated with the infinitesimal generators in Equation~\eqref{eq:lin_1} and Equation~\eqref{eq:lin_2}, given by
\begin{align}
X_1 &= y_1\partial_{y_1} + y_2\partial_{y_2},\\
X_2 &= y_2 \partial_{y_1} + y_1 \partial_{y_2},
\end{align}
found to generate symmetries of the linear model in Equation~\eqref{eq:lin}.

The 1-parameter groups $G_1$ and $G_2$ generated by $X_1^{(1)}$ and $X_2^{(1)}$, respectively, act on the total space $E$ according to
\begin{align}
\label{eqn:state_space_individual_action1}
\Gamma_{5,\epsilon} (t,y_1,y_2) & =  (t,e^{\epsilon}y_1,e^{\epsilon}y_2),\\
\Gamma_{6,\epsilon} (t,y_1,y_2) & =  (t,\cosh(\epsilon)y_1+\sinh(\epsilon)y_2,\sinh(\epsilon)y_1+\cosh(\epsilon)y_2),
\label{eqn:state_space_individual_action2}
\end{align}
and, consequently, the dimensions of the corresponding generic orbits are both $s_0 = 1$. Since the dimension is non-decreasing under prolongations and bounded from above by the group dimension, we can immediately conclude that $s_1=1$ for both $X_1^{(1)}$ and $X_2^{(1)}$. This implies that there are $\mu_1 = \mathrm{dim}\,J^{(1)} - s_1 = 5 - 1 = 4$
functionally independent first order differential invariants of each generator according to Equation~\eqref{eqn:functionally_independent_differential_invariants}. To compute these invariants, we need the explicit form of the prolonged generators which, in contrast to previous cases, are non-trivial and given by
\begin{align}
X_1^{(1)} & = y_1\partial_{y_1} + y_2\partial_{y_2} + y'_1\partial_{y'_1} + y'_2\partial_{y'_2},\\
X_2^{(1)} & = y_2\partial_{y_1} + y_1\partial_{y_2} + y'_2\partial_{y'_1} + y'_1\partial_{y'_2}.
\end{align}
Starting by considering the invariants of $X_1$ satisfying $X_1^{(1)}(I)=0$, we obtain the characteristic system
\begin{equation}
\frac{\text{d}y_1}{y_1} = \frac{\text{d}y_2}{y_2} = \frac{\text{d}y'_1}{y'_1} = \frac{\text{d}y'_2}{y'_2} = \text{d}s, \qquad \frac{dt}{ds}=0.
\end{equation}
Therefore, a complete set of first order differential invariants of $G_1^{(1)}$ is given by
\begin{equation}
I_1 = t, \qquad I_2 = \frac{y_2}{y_1},\qquad I_3 = \frac{y'_1}{y_1},\qquad I_4 = \frac{y'_2}{y_2},
\end{equation}
and the most general system admitting $X_1$ as a symmetry generator is
\begin{equation}
\label{eqn:general_system_X1}
y'_i = y_i F_i\left( t , \frac{y_2}{y_1} \right),
\end{equation}
for two arbitrary functions $F_1$ and $F_2$. In particular, the linear model in Equation~\eqref{eq:lin} corresponds to the choice $F_1(x_1,x_2) = 1+x_2$ and $F_2(x_1,x_2) = 1+x_2^{-1}$, whereas the choice $F_1(x_1,x_2) = a(x_1)+b(x_1)x_2$ and $F_2(x_1,x_2) = c(x_1)+d(x_1)x_2^{-1}$ shows that the most general linear model
\begin{align}
\label{eqn:generalized_linear_model}
\begin{split}
y'_1 & = a(t)y_1 + b(t)y_2,\\
y'_2 & = c(t)y_1 + d(t)y_2,
\end{split}
\end{align}
is also invariant under the symmetry generated by $X_1$. In fact, this is the only symmetry of the special case of the linear two state model in Equation~\eqref{eq:lin} that remains after the generalisation to the linear model in Equation~\eqref{eqn:generalized_linear_model}.

% --------------------------------------------------------------

Moving on to the generator $X_2^{(1)}$, the invariants are first integrals of the characteristic equations
\begin{equation}
\frac{\text{d}y_1}{y_2} = \frac{\text{d}y_2}{y_1} = \frac{\text{d}y'_1}{y'_2} = \frac{\text{d}y'_2}{y'_1} = \text{d}s, \qquad \frac{\text{d}t}{\text{d}s}=0,
\end{equation}
giving a complete set of first order differential invariants as
\begin{equation}
I_1 = t,\qquad I_2 = y_1^2-y_2^2,\qquad I_3 = \frac{y'_1+y'_2}{y_1+y_2},\qquad I_4 = \frac{y'_1-y'_2}{y_1-y_2},
\end{equation}
and the most general model invariant under $G_2^{(1)}$ is of the form
\begin{equation}
H_i\left( t , y_1^2-y_2^2 , \frac{y'_1+y'_2}{y_1+y_2} , \frac{y'_1-y'_2}{y_1-y_2} \right) = 0,\qquad i=1,2,
\end{equation}
according to Theorem~\ref{thm:symmetry_condition_invariants}. Solving these equations for the derivatives $y'_1$ and $y'_2$, we find the equivalent form that is formulated as a two state system of ODEs
\begin{align}
\label{eqn:general_system_X2}
\begin{split}
y'_1 & = y_1 F_1(t,y_1^2-y_2^2) + y_2 F_2(t,y_1^2-y_2^2),\\
y'_2 & = y_1 F_2(t,y_1^2-y_2^2) + y_2 F_1(t,y_1^2-y_2^2),
\end{split}
\end{align}
for two arbitrary functions $F_1$ and $F_2$. In particular, the special case of the two state linear model in Equation~\eqref{eq:lin} is recovered by the choice $F_1(x_1,x_2) = F_2(x_1,x_2) = 1$ for the arbitrary functions. 

This example demonstrates how a well-known class of models in mathematical biology, namely linear systems of ODEs, are connected to symmetry transformations acting on the state space. Moreover, an observation can be made about the most general two state system of ODEs that we obtained from the one-dimensional symmetry groups $G_1$ and $G_2$ generated by the infinitesimal generators $X_1$ and $X_2$. This observation is that the classes of ODEs in Equation~\eqref{eqn:general_system_X1} and Equation~\eqref{eqn:general_system_X2} are quite large in the sense there is a lot of room for choosing the arbitrary functions $F_1$ and $F_2$ in the reaction terms when designing a model from each class. In order to narrow down the possible choices of these reaction terms, we will now repeat this analysis for a higher-dimensional group that includes more than one infinitesimal generator. 

% --------------------------------------------------------------

\subsection{Higher-dimensional symmetry groups}

As we have seen in previous examples, requiring invariance under a 1-parameter group of symmetries restricts the admissible form of a model. Similarly, imposing invariance under higher-dimensional symmetry groups corresponds to simultaneously requiring the model to be invariant under all generators of $G$, further restricting the possible reaction terms. From a constructive model building perspective it is desirable to manifestly incorporate all known symmetries to eliminate structurally unfeasible models and increase biological interpretability.

For higher-dimensional symmetry groups, dimensional considerations become increasingly important in the computation of complete sets of differential invariants, which we will exemplify for the case of two dependent variables for the group $G$ generated by the Lie algebra $\mathfrak{g}_E=\mathrm{Span}(X_1,X_2)$ spanned by the infinitesimal generators of the linear model in Equation~\eqref{eq:lin} given by $X_1 = y_1\partial_{y_1}+y_2\partial_{y_2}$ and $X_2 = y_2\partial_{y_1}+y_1\partial_{y_2}$ considered in the previous example. Since $[X_1,X_2]=0$, the group $G$ is abelian and the action of a group element $\Gamma_g=\exp(\epsilon_1X_1+\epsilon_2X_2)$ on $E$ is given directly by Equations~\eqref{eqn:state_space_individual_action1} and~\eqref{eqn:state_space_individual_action2} as
\begin{equation}
\label{eqn:state_space_group_action}
\Gamma_g(t,y_1,y_2) = \left(t,e^{\epsilon_1}(\cosh(\epsilon_2)y_1+\sinh(\epsilon_2)y_2),e^{\epsilon_1}(\sinh(\epsilon_2)y_1+\cosh(\epsilon_2)y_2) \right).
\end{equation}
From the action it is clear that the dimension of a generic orbit of $G$ is $s_0=2=\mathrm{dim}\,G$, which, again, by the non-decreasing property of the dimension under prolongations implies that $s_1=2$. Consequently, the number of functionally independent first order differential invariants is $\mu_1 = \mathrm{dim}\,J^{(1)} - s_1 = 5 - 2 = 3$ according to Equation~\eqref{eqn:functionally_independent_differential_invariants}. Comparing this to the case of a single generator, we see that the number of functionally independent invariants is reduced by one, i.e.~from four to three invariants. Better still, due to the autonomy of this group, $I_1=t$ is trivially an invariant of $G$ which implies that the number of invariant combinations in state space that can appear in the reaction terms is reduced once more from three to two. This reduction is the manifestation of the requirement that a differential invariant $I$ of $G$ is simultaneously invariant with respect to each generator $X_1^{(1)}(I)=X_2^{(1)}(I)=0$.

From the previous example, we know that any function satisfying $X_1^{(1)}(I)=0$ can be written as $I=I(t,r_1,r_2,r_3)$ with
\begin{equation}
r_1 = \frac{y_2}{y_1},\qquad r_1 = \frac{y'_1}{y_1},\qquad r_3 = \frac{y'_2}{y_1},
\end{equation}
where we have made a different choice for the last differential invariant than above. Inserting the expression for $I$ into the remaining constraint yields
\begin{equation}
X_2^{(1)}(I) = (1-r_1^2)\frac{\partial I}{\partial r_1} + (r_3-r_1r_2)\frac{\partial I}{\partial r_2} + (r_2-r_1r_3)\frac{\partial I}{\partial r_3} = 0,
\end{equation}
for which the characteristic equations are
\begin{equation}
\frac{\text{d} r_1}{1-r_1^2} = \frac{\text{d} r_2}{r_3-r_1r_2} = \frac{\text{d} r_3}{r_2 -r_1r_3} = \text{d}s,\qquad \frac{\text{d}t}{\text{d}s} = 0.
\end{equation}
The first integrals of this system are given by
\begin{equation}
I_1 = t,\qquad I_2 = \frac{r_2+r_3}{1+r_1},\qquad I_3 = \frac{r_2-r_3}{1-r_1},
\end{equation}
and, in terms of the original jet space coordinates, a complete set of first order differential invariants is therefore
\begin{equation}
I_1 = t,\qquad I_2 = \frac{y'_1+y'_2}{y_1+y_2},\qquad I_3 = \frac{y'_1-y'_2}{y_1-y_2}.
\end{equation}
The most general form of a system of first order ODEs admitting the symmetry group $G$ is therefore, by an argument identical to that in the previous example, given by
\begin{align}
\label{eqn:general_system_X1_and_X2}
\begin{split}
y'_1 & = y_1 F_1(t) + y_2 F_2(t),\\
y'_2 & = y_1 F_2(t) + y_2 F_1(t),
\end{split}
\end{align}
for two arbitrary functions $F_1$ and $F_2$. Here, we note that the special case of the linear model in Equation~\eqref{eq:lin} with the autonomous reaction terms $\omega_1(t,y_1,y_2)=\omega_2(t,y_1,y_2)=y_1+y_2$ is recovered by setting $F_1(x_1)=F_2(x_1)=1$. In fact, if we add the time translation generator $\partial_t$ that is common to all autonomous models to the Lie algebra $\mathfrak{g}_E$ generating the group $G$, the time-dependence in the reaction terms of Equation~\eqref{eqn:general_system_X1_and_X2} would vanish, implying that the arbitrary functions in this case would be replaced by two constants, e.g.~$F_1(x_1)=C_1$ and $F_2(x_1)=C_2$. Moreover, by comparing the ODE system resulting from the two-dimensional group $G$ in Equation~\eqref{eqn:general_system_X1_and_X2} to the corresponding ODE systems in Equations~\eqref{eqn:general_system_X1} and~\eqref{eqn:general_system_X2} derived from the one-dimensional groups $G_1$ and $G_2$, respectively, we conclude that the admissible model structure is further restricted by imposing invariance under a larger symmetry group. This method of constructing ODE models from the invariants of a set of infinitesimal generators gives us a method for incorporating biological properties in the reaction terms of the models at hand. Based on this, we next propose a vision for how we can construct more realistic models of biological systems based on well-defined underlying principles.

% Model structure estimation
\section{Discovering biological mechanisms: estimating symmetries from experimental data}
\label{sec:symmetry_discovery}
\begin{figure}[htbp!]
\begin{center}
\includegraphics[width=0.8\textwidth]{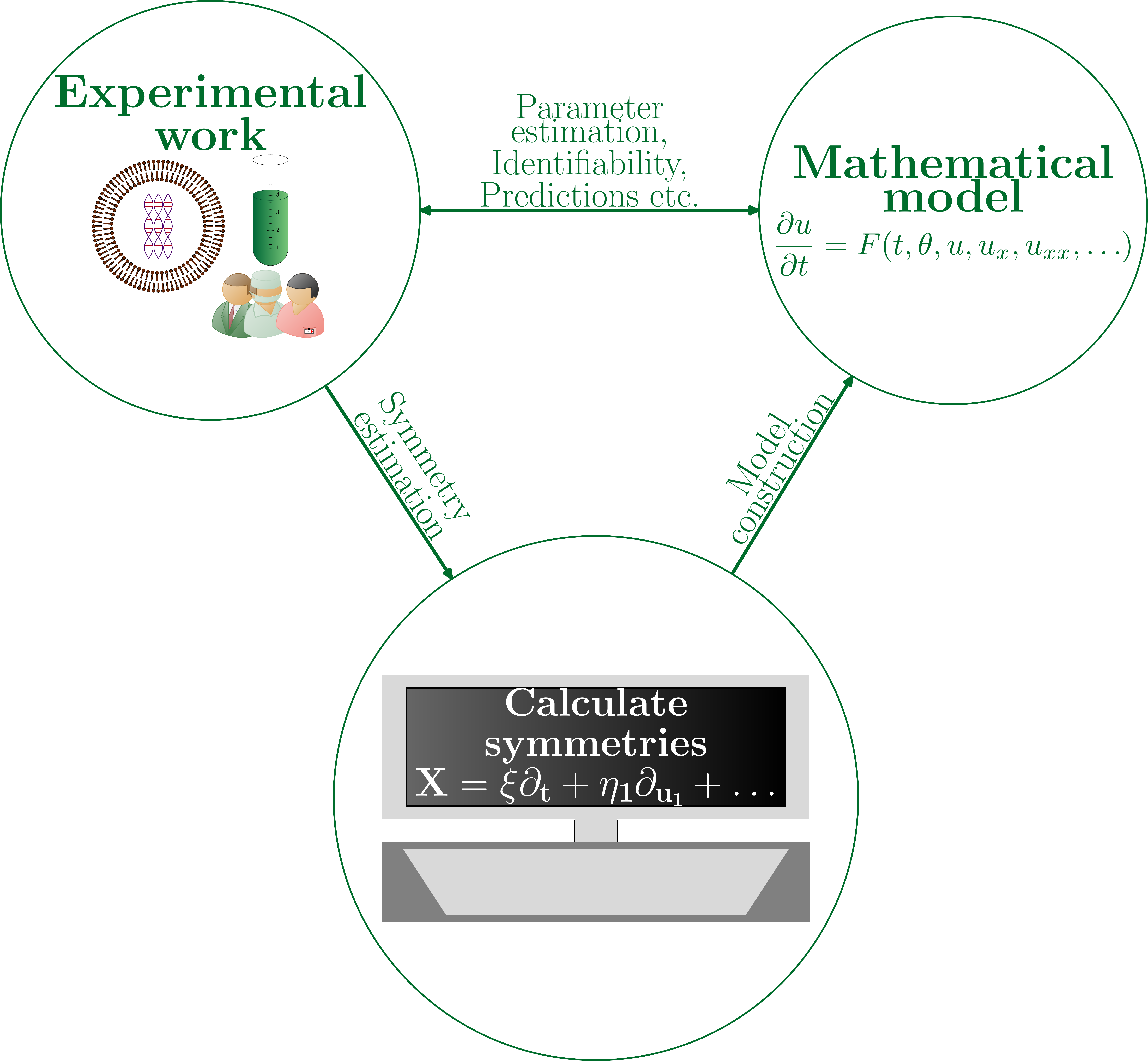}
\caption{The vision for symmetries in mathematical biology: model structure estimation. The idea of \emph{model structure estimation} is to estimate the model structure from experimental data based on the symmetries of the studied biological system. By building a framework for estimating the symmetries of a studied system from experimental data, it would be possible to construct candidate models using the invariants of the estimated symmetries. These models can then be used in an interdisciplinary framework in mathematical biology that combines experimental investigations with the techniques of parameter estimation, inference, identifiability analysis and model-based prediction. This figure has been produced by TikZ where its various parts are based on pictures produced by members of the TikZ community. In particular, the following sources of inspiration deserve a mention: the lipid vesicle by Henrik Skov Midtiby, the \textit{tikzpeople} package by Nils Fleischhacker and the computer by Elke Schubert.}
\label{fig:vision}
\end{center}
\end{figure}

In total, the symmetry methods presented in this work allow us to construct more realistic models of biological systems. As we saw in Section~\ref{sec:bioexamples}, the symmetries of biological systems encode properties such as the time-independence of autonomous models, the mass conservation of the SIR model and the energy conservation of the Lotka-Volterra model. Subsequently, in Section~\ref{sec:model_construction}, we demonstrated how to construct system of ODEs starting from a set of symmetries based on differential invariants. Thus, with a methodology for estimating symmetries from experimental data, we would be able to construct models based on the differential invariants of these estimated symmetries. We refer to this vision as \emph{Model structure estimation}, see Figure~\ref{fig:vision}.  

% ---------------------------------------------------------------------

% ---------------------------------------------------------------------

This approach has the potential for drastically improving our capacity to build mechanistically sound models as well as helping tackle one of the biggest obstacles in mathematical biology, namely that of model selection. Due to the complexity, and lack of knowledge about the fundamental properties, of biological systems, it is often possible to construct multiple candidate mechanistic models of a studied system. Oftentimes, the choice of a model for a given biological system is made by the modeller attempting to answer a \emph{model selection problem}: choose the candidate model that best fits the collected data. However, the fundamental, problem of model selection, which has been elegantly demonstrated in the context of cancer modelling~\cite{gerlee2013model}, is that multiple candidate models can fit the same data equally well. This implies that the candidate models are indistinguishable or, more importantly, that the underlying biological assumptions of these models cannot be differentiated based on the collected data. A further problem with the model selection approach is the implicit, yet fundamental assumption, that one of the candidate models is correct in the sense that it captures the underlying mechanisms. However, since models are necessarily simplifications, \emph{all} candidate models are incorrect in some way. In contrast, a theoretical approach of estimating the symmetries that are manifest in an observed system, in order to build a model based on the properties they encode, could in principle circumvent this problem.  

% Discussion
\section{Discussion}
\label{sec:discussion}
In this work, we have showcased the role of symmetries in the context of constructing and interpreting mechanistic models consisting of first order ODEs in mathematical biology. Based on the theory of symmetry methods for differential equations, we presented an algorithm for finding a particular class of symmetries of systems of ODEs with rational reaction terms along with an open source implementation of this algorithm that can be accessed at \url{https://github.com/JohannesBorgqvist/symSys_1st_ODEs}. Using our implementation of the algorithm, we calculated infinitesimal generators of symmetries of a number of well-known models in mathematical biology including the SIR and the Lotka--Volterra models. We then interpreted the meaning of the symmetries of these two models by deriving the corresponding differential invariants and from them three important properties: autonomy, mass conservation and energy conservation. We implemented the symmetry-based analysis in the reverse direction, that is, instead of deriving the symmetries of a particular model, we derived the most general class of models that has a particular symmetry. In particular, we derived the most general class of models associated with the symmetries we found previously using our algorithm and we demonstrated that the size of the class of constructed models is reduced as more symmetries are included in the construction phase. Lastly, we proposed the vision of \emph{model structure estimation} which entails estimating symmetries from data and thereafter constructing models using the differential invariants of these symmetries. In this way, we can automate the process of building biological properties into models and thereby construct more realistic models that capture the underlying mechanisms of the system at hand through their symmetries. 

Our algorithm provides a first step towards automating the calculations of symmetries. As symmetry methods view differential equations as geometrical objects, all the independent variables, such as the time and the dependent variables, are viewed as dimensions in a manifold. In practice, calculating symmetries entails solving a high-dimensional non-linear system of PDEs, such as the one in Equation~\eqref{eq:det_eq_practice}, and this strongly motivates the development of an automated or computer-assisted approach. We have used a set of projective ans\"atze together with a symbolic solver based on \textit{SymPy}~\cite{meurer2017sympy} which, in theory, captures a wide class of generators for systems of ODEs with rational reaction terms. Compared to previous implementations that use polynomial ans\"atze with constant coefficients~\cite{merkt2015higher}, our algorithm constitutes a generalisation. However, in practice our implementation is limited by the inefficiency of carrying out the symbolic calculations and we cannot currently even in principle use our approach to establish symmetries corresponding to non-polynomial infinitesimal generators. Thus, moving forward it will be critical to develop efficient algorithms for finding symmetries based on non-polynomial tangential ans\"atze. In addition, we must design tangential ans\"atze that are able to capture biological properties; this will entail the systematic calculation and interpretation of the symmetries of well-known biological models. 

We have demonstrated how well-known properties of the SIR and Lotka--Volterra models can be understood in terms of their symmetries. Specifically, we analysed the invariants of the symmetries of these models and saw that there are two types of invariants, classified by their corresponding symmetries. Firstly, if the symmetry is \emph{trivial} meaning that it maps points on one solution along \emph{the same solution curve}, then the corresponding invariant corresponds to \emph{conservation laws} such as \emph{mass conservation} in the case of the SIR model or \emph{energy conservation} in the context of the Lotka--Volterra model. Secondly, if the symmetry is \emph{non-trivial} meaning that it maps a solution to \emph{another} distinct solution then these invariants correspond to properties of the space of all solutions, such as the autonomy or time-independence of both the SIR and Lotka--Volterra models. By repeating this type of analysis for a large number of models in mathematical biology, we can characterise a certain biological or dynamical property by a symmetry and, in this way, we can ultimately construct a database of symmetries displayed by biological systems. Using this database, we can, on the one hand, wisely design the tangential ans\"atze used in the previously discussed algorithms for finding the symmetries and, on the other hand, use our knowledge to perform \emph{model structure estimation} using experimental data. 

We argue that the ultimate goal for symmetries in mathematical biology is to use them as the basis for the construction of mathematical models in which biological mechanisms are manifest. We propose that future research efforts should be directed towards developing a framework for estimating symmetries from experimental data and then constructing models from the differential invariants of these estimated symmetries (see Figure~\ref{fig:vision}). In this work, we showed how this latter part is done in practice by deriving classes of models starting from symmetries and, most importantly, we showed that the more symmetries we include in this construction phase the more precisely can we determine the class of models obeying these symmetries. The design of methods for estimating symmetries from experimental data remains an open and difficult problem. We can speculate how this might be done in the context of first order time-dependent ODEs where the data consists of time series. In order to estimate the symmetries of a given system, numerous time series with \emph{different} initial conditions are required. The objective is then to construct transformations that maps any time series to to another one in the data set. Such a transformation would constitute a symmetry of the data set and consequently a candidate for a symmetry of the underlying system. In practice, learning such transformations will require testing multiple different classes of functions as components of the infinitesimal generators and then use of the exponential map to retrieve the corresponding symmetry transformation.

As we have seen in the present paper, symmetries are immensely useful for understanding the properties of differential equations in the mechanistic modelling of biological systems. However, the scope of symmetries as a tool for modelling extends beyond this context. In a related approach, incorporating spatial symmetries of the input data in deep-learning models has produced remarkable results such as the discovery of protein structures starting from a sequence of amino acids~\cite{boomsma2017,jumper2021}. This work demonstrates the huge, further potential of symmetry methods for understanding and describing biological mechanisms across a host of temporal scales. 

\section*{Acknowledgements}

JB would like to thank the Wenner--Gren Foundation for a Research Fellowship and Linacre College, Oxford, for a Junior Research Fellowship. REB is a Royal Society Wolfson Research Merit Award holder.

% The bibliography
\clearpage
%\bibliographystyle{unsrt}
%\bibliography{symmetry_bib}

\end{document}